\documentclass[preprint,preprintnumbers,superscriptaddress,amsmath]{revtex4}
\usepackage{url}
\usepackage[dvips]{graphics}
\usepackage{epsfig}
\usepackage{psfrag}
\usepackage[psamsfonts]{amssymb}
\usepackage{amsmath}
\usepackage{indentfirst}
\usepackage{amssymb}
\usepackage{wrapfig}

\begin{document}

\title{Effects of Diversity and Procrastination in  Priority Queuing Theory: the Different Power Law Regimes}
\author{A. Saichev}
\affiliation{Department of Management, Technology and Economics,
ETH Zurich, Kreuzplatz 5, CH-8032 Zurich, Switzerland}
\affiliation{Mathematical Department,
Nizhny Novgorod State University, Gagarin prosp. 23,
Nizhny Novgorod, 603950, Russia}
\email{saichev@hotmail.com,dsornette@ethz.ch}

\author{D. Sornette}
\affiliation{Department of Management, Technology and Economics,
ETH Zurich, Kreuzplatz 5, CH-8032 Zurich, Switzerland}

\begin{abstract}

Empirical analysis show that, after the update of a browser,  the publication of
the vulnerability of a software, or the discovery of a cyber worm,
the fraction of computers still using the older version, or being not yet patched,
or exhibiting worm activity decays
as power laws $\sim 1/t^{\alpha}$ with $0 < \alpha \leq 1$ over time scales of years.
We present a simple model for  this persistence phenomenon framed within the
standard priority queuing theory, of a target task which has the lowest priority compared with all
other tasks that flow on the computer of an individual.
We identify a ``time deficit'' control parameter $\beta$
and a bifurcation to a regime where there is a non-zero probability for the target task
to never be completed.  The distribution of waiting time ${\cal T}$ till the completion
of the target task has the power law tail $\sim 1/t^{1/2}$, resulting from a first-passage
solution of an equivalent Wiener process. Taking into account a diversity of
time deficit parameters in a population of individuals, the power law tail is
changed into $1/t^\alpha$ with $\alpha\in(0.5,\infty)$, including the well-known case $1/t$.
We also study the effect of  ``procrastination'', defined
as the situation in which the target task may be postponed or delayed even after
the individual has solved all other pending tasks. This new regime
provides an explanation for even slower apparent decay and longer
persistence.

\end{abstract}

\date{\today}

\maketitle

\section{Introduction}

There is growing evidence that remarkably robust quantitative and sometimes
universal laws describe the behavior of humans in society.  Consider a
typical individual, who is subjected to a flow of information and of requested tasks,
in the presence of time, energy, regulatory, social and monetary constraints. Such an
individual will respond by a sequence of decisions and actions, which themselves
contribute to the flow of influences spreading to other individuals. A recently developed
approach to unravel at least a part of this complex ballet consists in quantifying
the waiting time distribution between triggering factor and response performed by
humans, which has been found in many situations to be a power law ${\rm pdf}(t) \sim 1/t^p$
with an exponent $p$ less than $2$. As a consequence, the mathematical expectation of the
waiting time between consecutive events  is infinite, which embodies the notion
of a very long persistence of past influences.
This power law behavior has been documented quantitatively for the
distribution of waiting times until an
email message is answered \cite{Eck}, for the time intervals between consecutive e-mails sent by a single user
and time delays for e-mail replies \cite{Barabasi_Nature05}, for the waiting time between receipt and
response in the correspondence of Darwin and of Einstein \cite{Oliveira_Bara},
and for the waiting times associated with other human activity patterns
which extend to web browsing, library visits and stock trading \cite{Vasquez_et_al_06}.

A related measure concerns the rate of activity following a shock, a perturbation,
an announcement and so on, that impacts a given social system. For instance, measures
of media coverage after a large geo-political event decay approximately
as a power law of time since the occurrence of the event
\cite{roehner2004news}. The rate of downloads of papers from a website
after a media coverage also follows a power law decay \cite{johansen2000internaut,johansen2001internaut}.
The rate of book sales following
an advertisement or large media exposition decays as a power law of the time since
that event \cite{sornette2004amazon,deschatres2005amazon}. The rate of
video views on YouTube decays also as a power law after peaks
associated with media exposure \cite{cranesorYouTube}. Ref.~\cite{sornette2005origins}
argues that many other systems are described by a similar behavior.

These two measures of human reactions, (i) distribution of waiting times
between triggering factor and response and (ii) rate of activity in response to a ``shock'',
 are related. This fact has been exemplified by the rate of
donations following the tsunami that occurred on December 26, 2004 \cite{CraneSchSor_donation}.
A donation associated with this event can be considered as a task that
was triggered (but not necessarily executed) on that day simultaneously
for a large population of potential donors. This task competes with many others associated with the jobs, private lives and other activities of each individual in the entire population. The social experiment provided by the tsunami illustrates
a general class of experiments in which the same ``singular task'' is presented at approximately the same time to all potential actors (here the donors), but the priority value of this singular task can be expected to be widely distributed among different individuals. Since the singular task has been initiated at nearly the same time for all individuals, the activity (number of donations) at a time $t$ after this initiation time is then simply equal to  $N \times {\rm pdf}(t)$, where $N$ is the number of individuals who will eventually act (donate) in the population and ${\rm pdf}(t)$ is the previously defined distribution of waiting times until a task is executed.

These observations have been rationalized by priority queuing models
that describe how the flow of tasks falling on (and/or self-created by) humans are executed using priority ranking
\cite{Barabasi_Nature05,Oliveira_Bara,Vasquez_et_al_06}.
Assuming that the average rate $\lambda$ of task arrivals is larger than the average rate $\mu$ for executing them, and
using a standard stochastic queuing model wherein tasks are
selected for execution on the basis of random continuous priority
values, Grinstein and Linsker derived the exact overall probability per
unit time, ${\rm pdf}(t)$, that a given task sits in the queue for a time $t$ before being executed  \cite{Grinstein1}:
\begin{equation}
{\rm pdf}(t) \sim \frac{1}{t^{3/2}}~,~~~{\rm for}~ \mu \leq \lambda~.
\label{eq:powerlawexptheorytroisdemis}
\end{equation}
Grinstein and Linsker showed that the distribution  (\ref{eq:powerlawexptheorytroisdemis}) is independent of the specific shape of the distribution of priority values among individuals \cite{Grinstein2}.
The value of the exponent $p=3/2$ is compatible with previously reported numerical
simulations \cite{Barabasi_Nature05,Oliveira_Bara,Vasquez_et_al_06} and with most
but not all of the empirical data.

Our present theoretical study can be considered both as a pedagogical
simplification and an extension of Grinstein and Linsker \cite{Grinstein2,Grinstein2},
with the goal of exploring different mechanisms explaining the deviations of the exponent $p$
for its canonical value $3/2$. In particular, we reveal below the fundamental statistical
origin of the power law (\ref{eq:powerlawexptheorytroisdemis}) with exponent $3/2$
as nothing but a first-passage problem of an underlying random walk \cite{Redner_first}.
The initial motivation for the present study came from
recent quantitative empirical studies \cite{Frei1,Frei2,Frei_powerlaw}
on the time decay of the use of outdated browsers on the Internet and of the remaining
detectable and surprisingly significant activity on the Internet
of the Blaster worm since 2003 up to present \cite{Jonathan,Frei_powerlaw}.
The activity in these systems reveals the equivalent of the survival
distribution of browsers or of computers which have not yet been updated or patched
(in the language of priority tasks, this is the fraction of all entities which have not
yet accomplished the task). These activities are found to decay as $\sim 1/t^\alpha$,
with $\alpha \equiv p-1$, where the exponent is one unit less than
for ${\rm pdf}(t)$ since it describes the decay of the complementary
cumulative (also known as the survival) distribution of entities that have not
yet acted. The new element is that $\alpha$ is found different from $1/2$,
sometimes smaller, while it is larger in other cases. Here, we ask what could be the
simplest explanations for such behaviors.

The structure of the paper is the following. Section \ref{jguuuurrw} presents
the model of a target task which has the lowest priority compared with all
other tasks that flow on the ``shoulders'' (or computer) of an individual. The distribution
of waiting time $\mathcal{T}$ till the completion of the target task is formulated as a first-passage
time of an approximately equivalent Wiener process with drift. There is
a control parameter, that we call the ``time deficit'' parameter $\beta$, which is
proportional to the drift of the associated Wiener process. It is
proportional to the difference between the average time $\langle \eta \rangle$
to complete a non-target task and the average time interval $\langle \tau \rangle$ between non-target task arrivals:
$\beta \propto \langle \eta \rangle - \langle \tau \rangle$. For small $\beta$'s, the probability
density function (pdf) $q(t)$ of $\mathcal{T}$ has a power law tail $1/t^{1+\alpha}$ with exponent
$\alpha=1/2$. Its corresponding complementary cumulative distribution $Q(t)$
exhibits a bifurcation as a function of $\beta$.
For $\beta <0$ but close to $0$, $Q(t) \sim 1/t^{\alpha}$
and tends to zero at long time. For $\beta >0$, $Q(t) \sim Q_{\infty} +C/t^{\alpha}$,
where $Q_{\infty}$ is the non-zero probability that the target task is never completed.
Section \ref{trjhbjqekfvkqef} extends the preceding analysis to a population of individuals
with different time deficit parameters $\beta$.  We distinguish between regular distributions
around $\beta=0$ and non-regular ones. For the former, the exponent $\alpha$ is
changed into the value $1$ by the effect of heterogeneity. For distributions of $\beta$ that allow for positive values,
the survival distribution $Q(t)$ exhibits again a non-zero asymptotic limit at large times.
For non-regular distributions of $\beta$, the exponent $\alpha$ is found to
be continuously tunable from $1/2$ to $+\infty$.
Section \ref{bwglvlw} introduces the mechanism of ``procrastination'', defined
as the situation in which the target task may be postponed or delayed even after
the individual has solved all other pending tasks. In the limit where
the procrastination inclination is large and the time deficit parameter
is close to zero, we find that the pdf $q(t)$ of $\mathcal{T}$
exhibits a new much slower power law tail $\sim 1/t^{1-\alpha}$, with $\alpha=1/2$
in the regular case.  The survival distribution $Q(t)$ is characterized
by a slow cross-over to the  asymptotic power law $\sim 1/t^{\alpha}$.
Section \ref{tghtgta} concludes.

\section{Model and standard solutions  \label{jguuuurrw}}

\subsection{Formulation in terms of a specific target task with the lowest priority}

Let us assume for definiteness, but without loss of generality, that the target task
is identified at the origin of time $t=0$. For concreteness, we will frame our
discussion by using the examples of the task of updating your browser
version on your computer to the newly available version. Another example
is the task of patching one of your softwares, after its vulnerability has been disclosed
and its patch has been made freely available. Our goal is to derive
the distribution of waiting times or, equivalently, the dependence with time
of the fraction of the population
that has not yet performed the task.

Starting with the classical theory of prioritized queue, we assume that the target task
has the lowest priority among all other user's tasks. In other words, the users
consider updating their browser or patching their softwares as doable only after
all their other tasks have been addressed. This captures the
casual empirical observation that computer users are often reluctant to interrupt their
work, social chatting and blogging, games and other activities on their computer for an update or patch
that often requires a complete shutdown and restart.

The time at which the target task
is performed is denoted $\mathcal{T}$: it corresponds to the time interval
over which the user has been busy doing other things. We therefore refer to it
as the ``busy time duration.'' By definition of $\mathcal{T}$,
for any $t\in(0,\mathcal{T})$, there are still other unsolved tasks that requires
the attention of the individual, while at the instant $t=\mathcal{T}$,
all tasks that arose earlier have been solved. In the present section, we assume
that, once freed of other preoccupations at time $\mathcal{T}$, the individual who has been presented
with the target task at $t=0$ will finally perform it immediately.
In section \ref{bwglvlw}, we investigate another situation in which, once free of other
constraints, the user nevertheless procrastinates. Then, new tasks may appear in the meantime,
leading to further delays in the
completion of the target task. This
procrastination mechanism leads to new slower decay laws and interesting
cross-over regimes. But, with the present assumption that the target
task is addressed as soon as the user is free of other tasks, we obtain that the
complementary cumulative distribution function $Q(t)$ of waiting times till the update
of the browser coincides with the probability that the busy time duration is
larger than the given instant $t$:
\begin{equation}\label{qwtthruprbusy}
Q(t) = \Pr\{\mathcal{T}>t\}~ .
\end{equation}

We now analyze in detail the components contributing to the busy time duration $\mathcal{T}$.
Consider first all the tasks which were present before $t=0$ when the new
target task was first presented to the individual, and which have not yet been completed.
Let us assume that a time $\eta_0$ is still need after $t=0$ to solve these tasks.
Then, in the scenario in which no new task occur, we have
\begin{equation}
Q(t) = \Pr\{\eta_0>t\} ~ .
\end{equation}
We consider now all the other scenarios in which new tasks may fall on the shoulders
of the individual after $t=0$. Specifically, let us assume that the number of such new tasks
grow with $t$ according to some staircase function $n(t)$, which increases by one unit at the
discrete occurrence times
\begin{equation}\label{tonen}
0<T_1 <T_2 \dots < T_{n(t)}<t~.
\end{equation}
We denote by $\eta_k$ the time needed by the individual to solve the $k$-th task.
Then, by definition of the  busy time duration $\mathcal{T}$, $Q(t)$ is given by the
probability of the chain of events $\bigcap_{k=0}^{n(t)} \mathcal{A}_k$,
\begin{equation}\label{achainformatht}
Q(t) = \Pr\{\mathcal{T}>t\} \equiv \Pr\left[\bigcap_{k=0}^{n(t)} \mathcal{A}_k\right]~ ,
\end{equation}
where
\begin{equation}\label{akwktndef}
\mathcal{A}_k = \eta_0+W(k) - T_{k+1} > 0
\end{equation}
and
\begin{equation}\label{wksum}
W(k) = \sum_{i=1}^k \eta_i~ , \qquad W(0) = 0~ .
\end{equation}
In (\ref{akwktndef}), it is understood that $T_{n(t)+1} \equiv t$

Expression (\ref{achainformatht}) shows that the
chain $\bigcap_{k=0}^{n(t)} \mathcal{A}_k$ of events $\mathcal{A}_k$ defined by (\ref{akwktndef})
determines $Q(t)$ completely. It is thus important to have a detailed understanding of it. First,
the event with index $0$ is nothing but
\begin{equation}
\mathcal{A}_0 \equiv \eta_0>T_1~,
\end{equation}
which corresponds to the scenario in which the time $\eta_0$ that the individual needs to solve all tasks stored up to $t=0$  is larger than the time $T_1$ at which the first new task appears after the time $t=0$ at which the target task has been assigned.
Since the individual is still solving other tasks, she cannot perform the target tasks before the
arrival of the first new task at $T_1$. The event
\begin{equation}
\mathcal{A}_1 \equiv \eta_0+\eta_1 > T_2
\end{equation}
represents the scenario in which the individual is still solving the tasks that were not yet finished before $t=0$
or the new task that appeared at time $T_1$ when the second task occurs at time
$T_2$. Again, the individual is busy until time $T_2$ and cannot address the target task. The set
of events $\mathcal{A}_k$ for all $k$'s up to $n(t)$ follow the same structure, so that the
individual has still had not time to address the target task at time $t < \mathcal{T}$.
Figure~1 illustrates this chain $\bigcap_{k=0}^{n(t)} \mathcal{A}_k$  of events.

\begin{quote}
\centerline{
\includegraphics[width=11cm]{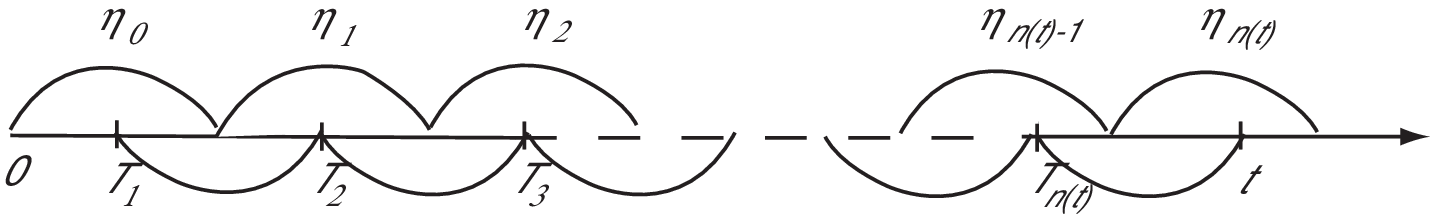}}
{\bf Fig.~1:} \small{Schematic illustration of the chain of events $\bigcap_{k=0}^{n(t)} \mathcal{A}_k$, leading to the occurrence of $\mathcal{T}>t$, where $\mathcal{T}$ is the  ``busy time duration'' until the target task is performed.}
\end{quote}

We now introduce the auxiliary stochastic process
\begin{equation}\label{vkdef}
V(k)= \sum_{i=1}^k (\eta_k- \tau_k)~ , \qquad V(0) = 0 ~ ,
\end{equation}
where
\begin{equation}
\tau_k = T_{k+1}-T_{k}
\end{equation}
is the time interval separating the occurrence of the $k$-th and $(k+1)$-th tasks arising after $t=0$.
It is also convenient to define
\begin{equation}\label{etazerenorm}
\eta_0' = \eta_0 - T_1~ ,
\end{equation}
as the time missing to complete all tasks stored up to $t=0$ when the first new task occurs at time $T_1>0$.
The case $\eta_0' \leq 0$ is excluded as it would correspond to the completion of the target task
at time $\eta_0$ before the arrival of the first task at time $T_1$. In this case, all subsequent tasks become irrelevant.
We also assume for simplicity that $\eta_0'$ is a fixed deterministic value (we will relax this
condition later on), so that
\begin{equation}\label{vketaprineq}
 \Pr\left[\bigcap_{k=0}^{n(t)} \mathcal{A}_k\right] = \Pr\{V(k)>-\eta_0' \}~,
\qquad \text{for any} ~~~ k\in(0,n(t))~ .
\end{equation}
The sought complementary distribution $Q(t)$ defined by expression \eqref{achainformatht} is therefore given by\begin{equation}\label{qtgenexpr}
Q(t) = \Pr\{V(k)>-\eta_0':k\in(0,n(t))\}~ .
\end{equation}
Figure~2 shows a typical realization of the stochastic process $V(k)$ defined by expression (\ref{vkdef}) over
a time interval in which the inequality $V(k)>-\eta_0'$ defining $Q(t)$ in (\ref{qtgenexpr}) holds.
\begin{quote}
\centerline{
\includegraphics[width=11cm]{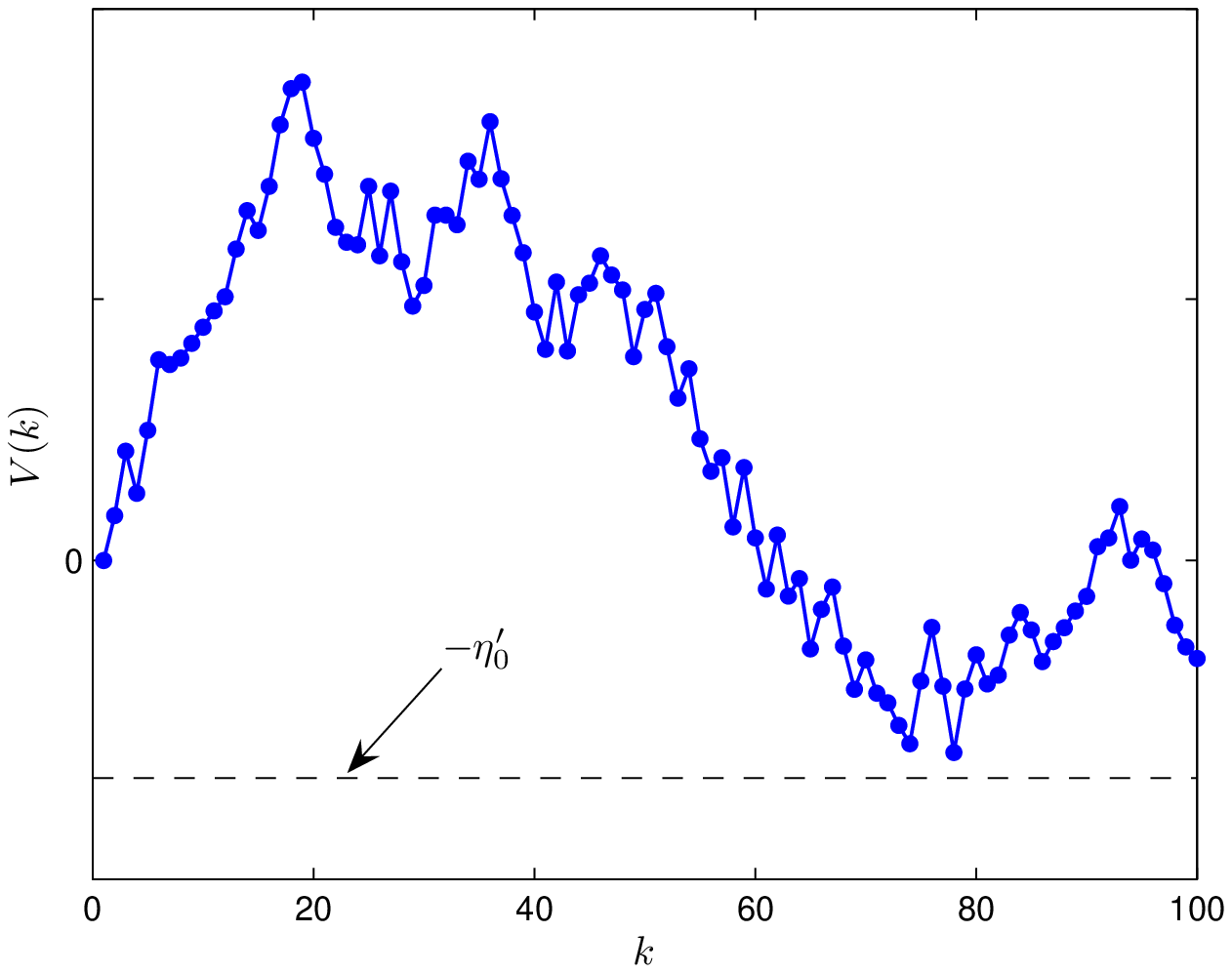}}
{\bf Fig.~2:} \small{Typical realization of the stochastic process $V(k)$ defined by expression (\ref{vkdef})
as a function of the discrete argument $k$ indexing the successive tasks appearing after $t=0$ at which
the target task has been initiated. This realization shown here obeys the inequality
$V(k)>-\eta_0'$ defining $Q(t)$ in (\ref{qtgenexpr}) over the whole time interval shown.}
\end{quote}

\subsection{Approximation in terms of a Wiener process (random walk) with drift \label{rbk'r}}

We assume that the two sequences $\{\eta_k\}$ and $\{\tau_k\}$ are made of  i.i.d random numbers,
with mean and variance respectively equal to  $\left<\eta\right>$, $\sigma_\eta^2$, and $\left<\tau\right>$, $\sigma_\tau^2$.

In order to obtain the asymptotical properties of $Q(t)$ given by \eqref{qtgenexpr} at large times
$t\gg \left<\tau\right>$, we apply the Law of Large Numbers (LLN) that
justifies replacing in \eqref{qtgenexpr} the random number $n(t)$ by its \emph{mean},
\begin{equation}\label{nstarrelat}
n(t)\approx \left<n(t)\right>=\theta \equiv t/\left<\tau\right>~ ,
\end{equation}
leading to the following asymptotically exact expression for $Q(t)$,
\begin{equation}\label{qtgeomprobinterpr}
Q(t) \approx \Pr\{V(k) > -\eta_0':k\in[0,\theta]\}~ .
\end{equation}

For large times, and thus large $\theta$'s, the Central Limit Theorem (CLT) ensures that the process $V(k)$ defined by \eqref{vkdef} can be interpreted as the discrete version of a Wiener process with drift, with the following mean value and variance
\begin{equation}\label{meanvarwienvn}
\left<V(k)\right> = k \cdot \left(\left<\eta\right> - \left<\tau\right>\right) ~ , \quad \text{Var}[V(k)] = \sigma^2 \cdot k ~ , \quad \sigma^2 = \sigma^2_\eta+ \sigma^2_\tau~ .
\end{equation}
Moreover, for $\theta\gg 1$, we can replace $V(k)$ by its continuous limit, in the form of the standard Wiener process with drift, depending on the \emph{continuous} argument $k$. Accordingly, the probability density function (pdf) $f(v;k)$ of the continuous stochastic process $V(k)$ satisfies the diffusion equation
\begin{equation}\label{difeqfvm}
{\partial f(v;k) \over \partial k} + (\left<\eta\right>-\left<\tau\right>) {\partial f(v;k) \over \partial v} = {\sigma^2 \over 2} {\partial^2 f(v;k) \over \partial v^2}~ ,
\end{equation}
supplemented by the initial condition
\begin{equation}\label{incondfm}
f(k;0) = \delta(v)~ .
\end{equation}
Recall that $f(v;k) dv$ is the probability to find $V(k)$ between $v$ and $v+dv$ at ``time'' $k$.

From the theory of Wiener processes, and within the asymptotically exact continuous limit just described,
it follows that the probability given \eqref{qtgeomprobinterpr} is given by
\begin{equation}\label{qtn4}
Q(t) = \int_{-\eta_0'}^\infty f(v;\theta|\eta_0') dv~ ,
\end{equation}
where $f(v;k|\eta_0')$ is the solution of the diffusion equation \eqref{difeqfvm} satisfying the initial condition \eqref{incondfm} and the additional absorbing boundary condition
\begin{equation}\label{fvn}
f(v;k|\eta_0')\big|_{v=-\eta'_0} = 0~ .
\end{equation}

The solution of the initial-boundary problem \eqref{difeqfvm}, \eqref{incondfm}, \eqref{fvn} is
\begin{equation}
\begin{array}{c}
f(v;k|\eta_0') = g(v-(\left<\eta\right>-\left<\tau\right>) k; k)-
\\[4mm] \displaystyle
\exp\left(-{2 (\left<\eta\right>-\left<\tau\right>) \eta_0' \over \sigma^2} \right) g(v- (\left<\eta\right>-\left<\tau\right>) k + 2 \eta_0';k)~,
\end{array}
\label{fvnboundsol}
\end{equation}
where
\begin{equation}
g(v;k) = {1 \over \sqrt{2 \pi k} \sigma} \exp\left(- {v^2 \over 2 \sigma^2 k}\right)~ .
\end{equation}
Substituting expression \eqref{fvnboundsol} into \eqref{qtn4} yields
\begin{equation}\label{qnt5}
Q(t) \approx {1 \over 2} \left(1+\text{erf}\left({\gamma + \delta \theta \over \sqrt{2\theta}}\right) - e^{-2 \delta \gamma} \text{erfc}\left({\gamma - \delta \theta \over \sqrt{2\theta}}\right)\right)~ ,
\end{equation}
with the following notations
\begin{equation}\label{taudelnot}
\gamma = {\eta_0' \over \sigma_\eta}~ , \qquad \delta =
{\left<\eta\right>-\left<\tau\right> \over \sigma}~ , \qquad \theta = {t \over \left<\tau\right>} ~.
\end{equation}
The corresponding pdf is
\begin{equation}\label{qpdftheta}
q(t) \equiv - {d Q(t) \over dt} =  {1 \over \left<\tau\right> } {\gamma \over \sqrt{2\pi} \theta^{3/2}} \exp\left(-{(\delta \theta + \gamma)^2 \over 2 \theta}\right)~ .
\end{equation}

In the analysis that follows, the key role played by the parameter $\delta$ defined in \eqref{taudelnot} warrants
further interpretation. Let us assume that the occurrence of new tasks at
the times $\{T_k\}$ defined by (\ref{tonen}) is a Poisson flow with rate $\lambda$.
Similarly, we assume that the completion of tasks at the times $\{W(k)\}$
defined by (\ref{wksum}) is also a Poissonian flow with rate $\mu$. We thus have
\begin{equation}
\left<\tau\right> = \sigma_\tau ={1 \over \lambda}~ , \qquad \left<\eta\right> = \sigma_\eta ={1 \over \mu}~ .
\end{equation}
$\left<\tau\right>$ is the mean time between arriving tasks and $\left<\eta\right>$ is the mean
completion time of the tasks.
Accordingly, the parameter $\delta$ is equal to
\begin{equation}\label{deltapoisson}
\delta = {\varepsilon -1 \over \sqrt{\varepsilon^2 +1}}~ , \qquad \text{where} \qquad \varepsilon = \lambda \left<\eta\right> = {\lambda \over \mu} ~ .
\end{equation}
When
\begin{equation}\label{exbalcond}
\left<\tau\right> = \left<\eta\right> \qquad \Rightarrow \qquad \delta = 0
\qquad \Rightarrow \qquad \varepsilon=1 ~ ,
\end{equation}
the rate of new task arrivals is equal to the rate of solving them. This  \emph{balanced} condition
corresponds to a zero drift in the associated Wiener process, and plays a crucial role
in the generation of power laws in the distribution of waiting times.
When positive, the parameter $\delta$ quantifies the ``time deficit'' that is missing on average per task
in order for the individual to finally be able to complete the target task. For a negative
time deficit ($\delta<0)$, the target task is almost surely performed in finite time, as we show below.

It is instructive to analyze separately the complementary cumulative distribution $Q(t)$ given by \eqref{qnt5} and its corresponding pdf $q(t)$ given by \eqref{qpdftheta} of the waiting time for the target task to be done (browser upgrade or software patched). We will discuss different situations in which $Q(t)$ and $q(t)$ are described by asymptotic
power laws
\begin{equation}\label{kaqqpowls}
Q(t) \sim Q_\infty + \theta^{-\alpha} \qquad \iff \qquad q(t) \sim \theta^{-\alpha-1}~ ,
\end{equation}
where $Q_\infty$ may be non-zero in some interesting cases to be discussed below.

\subsection{Derivation of the power law pdf $q(t)$ of waiting times till the completion of the target task}

We first rewrite the pdf $q(t)$ given by \eqref{qpdftheta} in a form more convenient for its analysis.
For this, we notice that $q(t)$ is controlled by two characteristic scales
\begin{equation}\label{thdthgdef}
\theta_\gamma = {\gamma^2 \over 2} ~ , \qquad \theta_\delta = {2 \over \delta^2}~ .
\end{equation}
For definiteness, consistent with our previous assumption that $\eta_0'$ is constant, we
take $\gamma$ (i.e. $\theta_\gamma$) to be constant. We then explore the behavior of
$q(t)$ for different values of $\delta$ (i.e. $\theta_\delta$). It is convenient to introduce the new variable
\begin{equation}\label{rhodef}
\rho = {\theta \over \theta_\gamma} = {2\theta \over \gamma^2 }
\end{equation}
and the parameter
\begin{equation}\label{epsdef}
\beta = \sqrt{{\theta_\gamma \over \theta_\delta}} = {1 \over 2} \gamma \delta~ .
\end{equation}
The dimensionless pdf
\begin{equation}
\kappa(\rho;\beta) ={\gamma^2 \over 2} \sqrt{\pi}  \left<\tau\right> q(t)
\end{equation}
takes the form
\begin{equation}\label{kaprhoexp}
\kappa(\rho;\beta) = {1 \over \rho^{3/2}} \exp\left(- {(\beta \rho+1)^2 \over \rho} \right)~.
\end{equation}

The dependence of $\kappa(\rho;\beta)$ as a function of $\rho$ is qualitatively different for $\beta\ll 1$ ($\theta_\gamma \ll \theta_\delta$) and for $\beta\gtrsim 1$ ($\theta_\gamma \gtrsim \theta_\delta$). For $\beta\gtrsim 1$, $\kappa(\rho;\beta)$
does not exhibit any power law asymptotic, not even in an intermediate domain of $\rho$. In contrast, for
\begin{equation}\label{betalessone}
\beta \ll 1 \qquad \iff \qquad {1 \over 2} \gamma \delta \ll 1 ~,
\end{equation}
the pdf $\kappa(\rho;\beta)$ possesses the intermediate power asymptotic
\begin{equation}\label{kaprhointpowasymp}
\kappa(\rho;\beta) \approx \rho^{-3/2} ~ , \qquad 1 \lesssim \rho \lesssim \beta^{-2} ~ ,
\end{equation}
which is replaced, for larger $\rho$, by the exponential decay
\begin{equation}\label{kaprhoexpasymp}
\kappa(\rho;\beta) \approx \rho^{-3/2} e^{-\beta^2 \rho}~ .
\end{equation}
Note that the function $\kappa(\rho;\beta)$ is nothing but the pdf of
first return to the absorbing boundary condition defined in (\ref{fvn}) (the target task is performed)
of the Wiener process with drift with the characteristics (\ref{meanvarwienvn}) \cite{Redner_first}.
In the balanced case \eqref{exbalcond}
$$
\delta = 0 \qquad \Rightarrow \qquad \beta^{-2} = \infty~ ,
$$
the power law \eqref{kaprhointpowasymp} holds for any $\rho\gtrsim 1$.

Figure~3 illustrates the cross-over of $\kappa(\rho;\beta)$ from the intermediate power law asymptotic (\ref{kaprhointpowasymp}) and the exponential tail (\ref{kaprhoexpasymp}) for three increasing values of the normalized
drift parameter $\beta$.

\begin{quote}
\centerline{
\includegraphics[width=11cm]{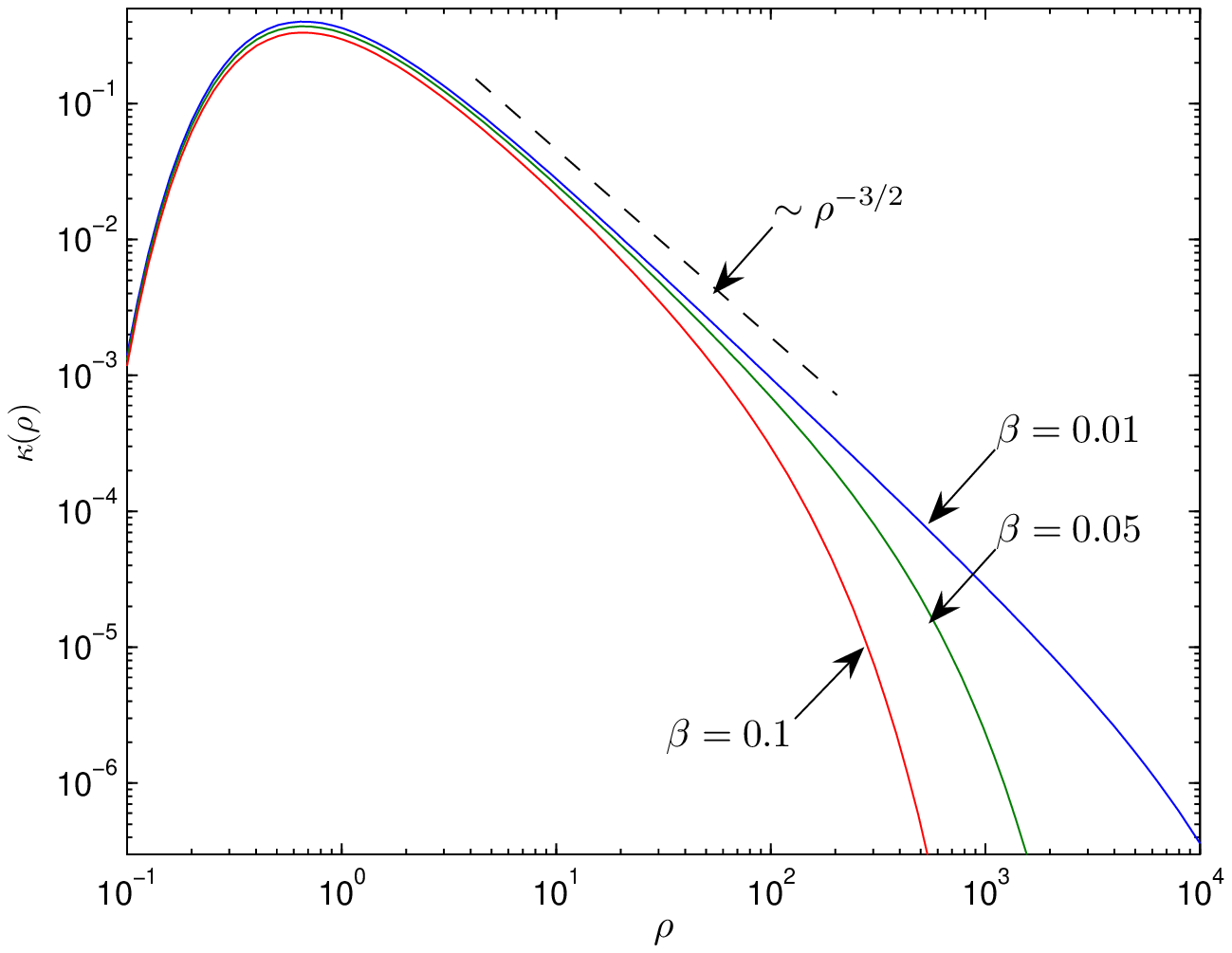}}
{\bf Fig.~3:} \small{Dependence of $\kappa(\rho;\beta)$ given by \eqref{kaprhoexp}
as a function of $\rho$ for three values of the normalized drift parameter $\beta=0.01; 0.05; 0.1$. The dashed straight line correspond to the pure power law $\sim \rho^{-3/2}$. The larger $\beta$ is, the narrower is interval in $\rho$ for which the intermediate power asymptotic holds.}
\end{quote}

\subsection{Derivation of the survival distribution $Q(t)$ of waiting times till the completion of the target task \label{thbbns}}

The complementary cumulative distribution $Q(t)$ given by \eqref{qnt5} can be rewritten in the
following form, which is more convenient for the analysis of its asymptotic behavior:
\begin{equation}\label{qcapthrurho}
Q(t) \equiv \mathcal{Q}(\rho;\beta) = {1 \over 2} \left( 1 + \text{erf}\left({1+\beta \rho \over \sqrt{\rho}}\right) - e^{-4\beta}\, \text{erfc}\left({1-\beta \rho \over \sqrt{\rho}}\right) \right)~ .
\end{equation}
This allows us to show that
the function $\mathcal{Q}(\rho;\beta)$ has a qualitatively different behavior at $\rho\to\infty$ for $\beta> 0$ and for $\beta< 0$.
This can be seen from the corresponding limits of $\mathcal{Q}(\rho;\beta)$:
\begin{equation}\label{qliminf}
\lim_{\rho\to\infty} \mathcal{Q}(\rho;\beta) =
\begin{cases}
Q_\infty(\beta) ~ , & \beta> 0~ , \\
0 ~ , & \beta < 0~ .
\end{cases}
\qquad Q_\infty(\beta) = 1-e^{-4 \beta}~ .
\end{equation}
For $\beta>0$, $Q(t)$ tends to a strictly \emph{positive limit} $Q_\infty(\beta)>0$ as $\rho \to +\infty$. For $\beta<0$,
$\mathcal{Q}(\rho;\beta)$ tends to zero as $\rho \to +\infty$. These two limits have
the following simple probabilistic interpretations.
For $\beta>0$, $ \left<\eta\right> > \left<\tau\right>$: the average time needed to complete a task
is larger than the average inter-arrival times between new incoming tasks. As a consequence,
there is strictly positive probability $Q_\infty(\beta)>0$ that the target task
will never be completed. In the language of the drifting Wiener process with
characteristics (\ref{meanvarwienvn}), in the presence of a positive drift, there is a finite
probability $Q_\infty(\beta)$ for the Wiener process to escape to infinity (the target
task is never completed), and a probability $1-Q_\infty(\beta)$ for
being captured at the absorbing boundary defined in (\ref{fvn}) (the target task is performed).
In contrast, for $\beta<0$, $\left<\eta\right> < \left<\tau\right>$, and the individual will
almost surely complete the target task (update her browser or patch her software) in finite time.
Thus, for $\beta<0$, $Q_\infty(\beta) = 0$.
Crossing the value $\beta=0$ is analogous to a phase transition or bifurcation
characterized by the order parameter $Q_\infty(\beta)$ varying
as a function of the control parameter $\beta$: for $\beta <0$, the order
parameter is zero and it bifurcates to a non-zero value for $\beta>0$.

Equation \eqref{qcapthrurho} expressed for the balanced case $\beta=0$ yields the following power law asymptotic
\begin{equation}\label{qbetouasymp}
\mathcal{Q}(\rho;0) \approx {2 \over \sqrt{\pi\rho}}~ , \qquad \rho\gtrsim 1~ ,
\end{equation}
corresponding to power law exponent $\alpha=1/2$, as defined in (\ref{kaqqpowls}).
For $|\beta|\gtrsim 1$, $\mathcal{Q}(\rho;\beta)$ does not exhibit any power law asymptotic,
not even in an intermediate domain of $\rho$.
For $|\beta|\ll 1$, the power law \eqref{qbetouasymp} holds as an intermediate asymptotic
in the interval
\begin{equation}\label{powaswithin}
1\lesssim\rho\lesssim\beta^{-2}~ .
\end{equation}
For $\rho\gg\beta^{-2}$, the intermediate power law asymptotic \eqref{qbetouasymp} crosses over
to an exponential decay converging to $0$ for $\beta<0$ or to $Q_\infty(\beta)>0$ given
in (\ref{qliminf}) for $\beta>0$:
\begin{equation}\label{qcompasymp}
\mathcal{Q}(\rho;\beta) \approx {1 \over \beta^2 \sqrt{\pi} \rho^{3/2}} e^{-2 \beta - \beta^2 \rho} +
\begin{cases}
Q_\infty(\beta)>0~ , & \beta> 0~ , \\
0 ~ , & \beta< 0 ~ .
\end{cases}
\end{equation}
Figure~4 plots the dependence of $\mathcal{Q}(\rho;\beta)$ as a function of $\rho$ for, $\beta=\pm 0.001$,
which illustrates the qualitatively different behavior of $\mathcal{Q}(\rho;\beta)$ for $\beta>0$ and for $\beta<0$.

\begin{quote}
\centerline{
\includegraphics[width=11cm]{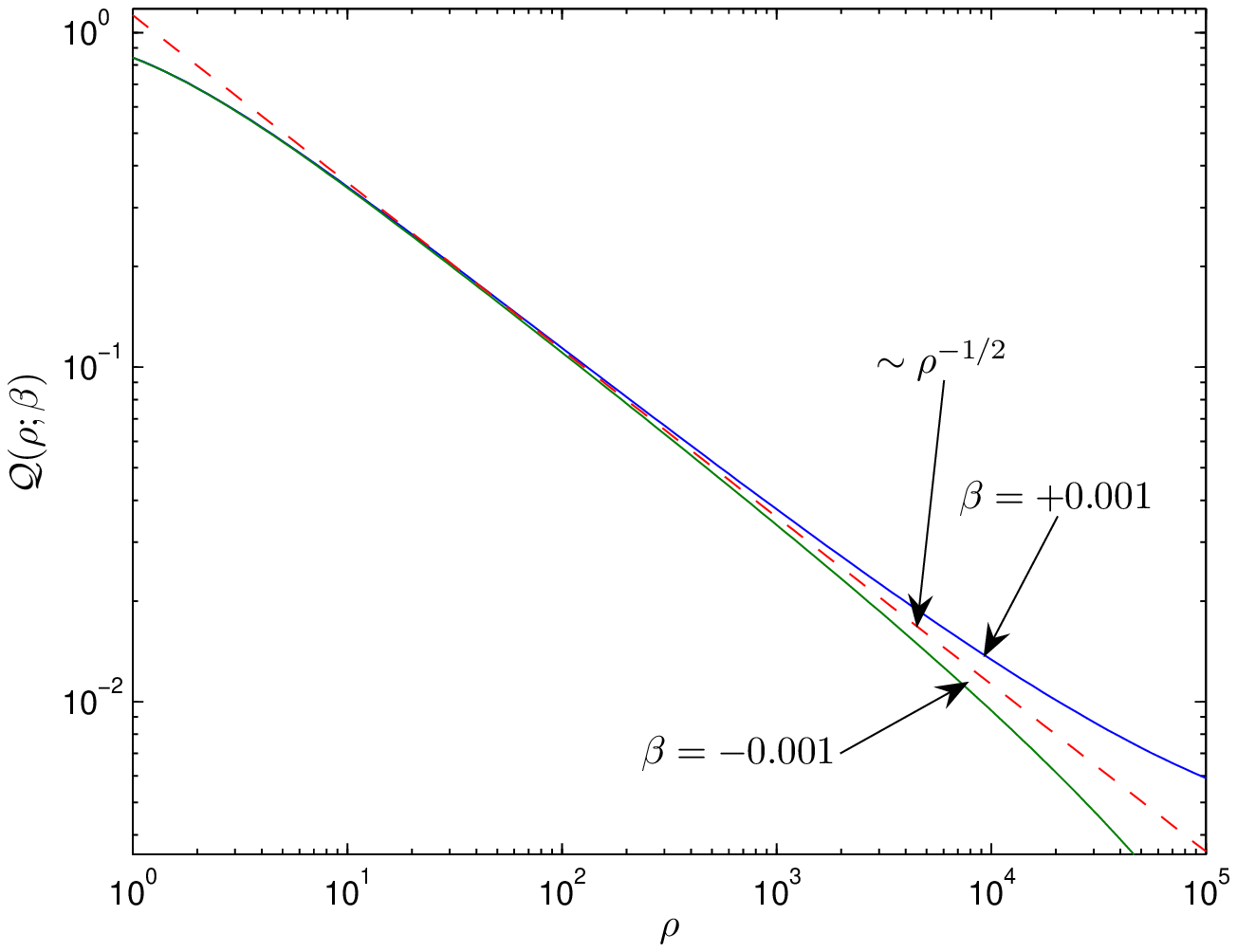}}
{\bf Fig.~4:} \small{Complementary cumulative distribution  $\mathcal{Q}(\rho;\beta)$ for $\beta=\pm 10^{-3}$ as a function of the normalized waiting time $\rho$ for completing the target task, demonstrating the qualitatively different asymptotic behavior of $\mathcal{Q}(\rho;\beta)$ for $\beta>0$ and for $\beta<0$. The dashed straight line shows the power asymptotic \eqref{qbetouasymp} corresponding to the balanced case $\beta=0$.}
\end{quote}

\section{Distributions of the time deficit parameter leading to different power law exponents and regimes \label{trjhbjqekfvkqef}}

\subsection{Regular distribution of the normalized time deficit parameter $\beta$ around the origin \label{yhyth3ya}}

\subsubsection{Qualitative justification of the form of the normalized time deficit parameter $\beta$ \label{gbja'awqfkv}}

As shown in the previous section, the proximity to the balance condition \eqref{exbalcond}
is essential for the power laws (\ref{kaprhointpowasymp})
or (\ref{qbetouasymp}) to hold over an intermediate asymptotic region
sufficiently large to be observable (at least over one to two decades in time).

In the present theory, the time deficit parameter $\delta$ or equivalently its normalized version $\beta$ are exogenously given.
In reality, $\delta$ ($\beta$) embodies the interplay between the subtle processes of task formation,
of prioritization and the efforts undertaken to solve them, that each individual adjusts continuously.
We conjecture that users of browsers and of softwares adapt approximately but not exactly, of course,
to the balance condition \eqref{exbalcond}. This is done for instance by being selective among the
flow of tasks (by deleting needless incoming emails or ignoring some superfluous tasks), and/or
by adapting the time allocated to solving tasks, so that
the mean time $\left<\eta\right>$ needed  to solve a given problem is approximately equal to the mean
time interval $\left<\tau\right>$ between subsequent arriving tasks. If this was not the case,
the individual would in general adjust the allocation of her time. Suppose for instance that
$\left<\eta\right> >\left<\tau\right>$. In this case, the individual is not able to face the flow
of incoming tasks and a boundless number of tasks piles up, suggesting a non-sustainable regime
either for the computer or its user. In the opposite case $\left<\eta\right> < \left<\tau\right>$,
the individual sits idle a significant fraction of her time. By enlarging the definition of what is
meant by ``task'' to include other activities, including the recreational activities that arguably
constitute a significant part of the utility or pleasure driving
individuals, it is clear that the case $\left<\eta\right> < \left<\tau\right>$ is not realistic
as a sustained regime. We also conjecture that the adjustment process leading
to the convergence of $\left<\tau\right>$ towards $\left<\eta\right>$ and vice-versa
may describe the general problem of task flow versus their solutions, beyond
the specific problem of browser update and software patching discussed here, to
encompass the general balance of human activities. We thus believe that the
results presented here are of broader interest and may help understand
the general statistical properties of the time allocation of humans.

Following these arguments, while individuals can be expected to adjust towards the balance
condition \eqref{exbalcond}, it is unlikely that all humans will do so accurately.  Indeed,
in many systems subjected to noise in which
state-dependent control actions are performed, the control parameter never settles
but continues to fluctuate around the target parameter \cite{Cabrera_Milton04,Eurich_Pawelzik05,Eurich_Pawelzik07}.
Therefore, we propose that the population
of browser and software users can be described by a distribution of time deficit parameters $\delta$ ($\beta$)
which is centered on $0$. We thus take into account the mentioned fluctuations by considering
$\delta$ as random variable with some pdf $\phi(\delta)$. The idealized balance condition $\delta=0$
is replaced by the more realistic \emph{mean balance condition}:
\begin{equation}\label{meanbalancecond}
\left<\delta\right> = \int_{-\infty}^\infty \delta \phi(\delta) d\delta = 0~ .
\end{equation}
Assuming that $\gamma$ is some deterministic constant as done above,
 then the mean balance condition  \eqref{meanbalancecond} is equivalent to
\begin{equation}\label{betabalancecond}
\left<\beta\right> = 0~ .
\end{equation}

\subsubsection{Derivation of the pdf $q(t)$ for a Gaussian distribution of the normalized time deficit parameter $\beta$}

In order to obtain concrete quantitative predictions, let us first consider that the pdf $\psi(\beta)$ of the random variable $\beta$
is a Gaussian law centered on $0$ and with standard deviation $\beta_0$:
\begin{equation}\label{gaussbeta}
\psi(\beta;\beta_0) = {1 \over \sqrt{2\pi} \beta_0} \exp\left(-{\beta^2 \over 2\beta_0^2} \right) ~ .
\end{equation}
The idealized balance condition $\beta=0$ is recovered in the limit $\beta_0 \to 0$ for
which the Gaussian pdf (\ref{gaussbeta}) tends to the Dirac function
$\psi(\beta) = \delta(\beta)$. With the choice (\ref{gaussbeta}), the mean balance condition (\ref{betabalancecond}) holds
by construction.

Amazingly, it turns out that the fluctuations of the parameter $\beta$,  that satisfy the mean balance condition $\left<\beta\right> = 0$, drastically change the asymptotic form of the distribution $Q(t)$ and its associated pdf $q(t)$, compared with
the idealized case in which the balance condition $\beta=0$ holds exactly for each individual at all times.
Let us first analyze this effect on the pdf of waiting times for the completion of the target task, which is now
given, in its normalized version by the weighted average with respect to $\beta$ of
$\kappa(\rho;\beta)$ given by \eqref{kaprhoexp}:
\begin{equation}\label{kapbarexpr}
\bar{\kappa}(\rho) = \int_{-\infty}^\infty \kappa(\rho;\beta) \psi(\beta) d\beta~ .
\end{equation}
Using the Gaussian distribution \eqref{gaussbeta}, this yields
\begin{equation}\label{avepdfkapexpr}
\bar{\kappa}(\rho;\beta_0) = {1 \over \rho^{3/2} \sqrt{1+2 \beta_0^2 \, \rho}}~ \exp\left(- {2\beta_0^2 \over \rho(1 + 2 \beta^2_0 \rho)} \right)~ ,
\end{equation}
which has the following asymptotic behavior
\begin{equation}\label{avepdfkappowas}
\bar{\kappa}(\rho;\beta_0) \approx {1 \over \sqrt{2\pi}\beta_0}~ {1 \over \rho^2} \sim \rho^{-2} ~ , \qquad  \rho\gtrsim \beta_0^{-2}~ .
\end{equation}
Thus,  the exponent $\alpha$, defined in
(\ref{kaqqpowls}), changes from the value $\alpha=1/2$ for the idealized balance condition into $\alpha=1$ in the presence of fluctuations of the time deficit parameter from individual to individual and/or as a function of time.
The mechanism acting here is similar to the mechanism of  ``sweeping of an instability'' \cite{SweepSor94}, since the
presence of a distribution of time deficit parameters around the balance
condition $\beta=0$ indeed amounts to sweeping the control parameter $\beta$ over its bifurcation point
defined in subsection \ref{thbbns}. We stress that this ``renormalization'' of the exponent $\alpha$ from the
value $1/2$ to $1$ is not sensitive to the details of the shape (\ref{gaussbeta}) of the distribution
of the time deficit parameter $\beta$. The single essential feature is that $\psi(\beta;\beta_0)$
goes to a constant for $\beta \to 0$. For any distribution $\psi(\beta;\beta_0)$ having
this property of going to a non-zero constant as $\beta \to 0$, the asymptotic tail (\ref{avepdfkappowas}) holds.
We will discuss in subsection \ref{tuhfvnghihd} variations to this
conditions and derive the corresponding changes in the exponent $\alpha$.

For the specific form (\ref{gaussbeta}), the above asymptotic result (\ref{avepdfkappowas})
can be made more accurate as follows.
For $\beta_0\gtrsim 1$,  $\bar{\kappa}(\rho;\beta)$ presents the unique power law regime
(\ref{avepdfkappowas}) with exponent $\alpha=1$.
For $\beta_0\ll 1$, there is an additional intermediate power law asymptotic with exponent $\alpha=1/2$
in the interval (analogous to \eqref{powaswithin})
\begin{equation}\label{powaswithinbou}
1\lesssim\rho\lesssim\beta_0^{-2}~ ,
\end{equation}
which is followed beyond the crossover point $\rho_*\approx \beta_0^{-2}$ by the power law (\ref{avepdfkappowas})  with exponent $\alpha=1$. Figure~5 shows the dependence of $\bar{\kappa}(\rho;\beta)$ as a function of $\rho$ given by
expression \eqref{avepdfkapexpr}, for three different values of $\beta$ that illustrate the intermediate asymptotic
with $\alpha=1/2$ and the tail asymptotic with $\alpha=1$.

\begin{quote}
\centerline{
\includegraphics[width=11cm]{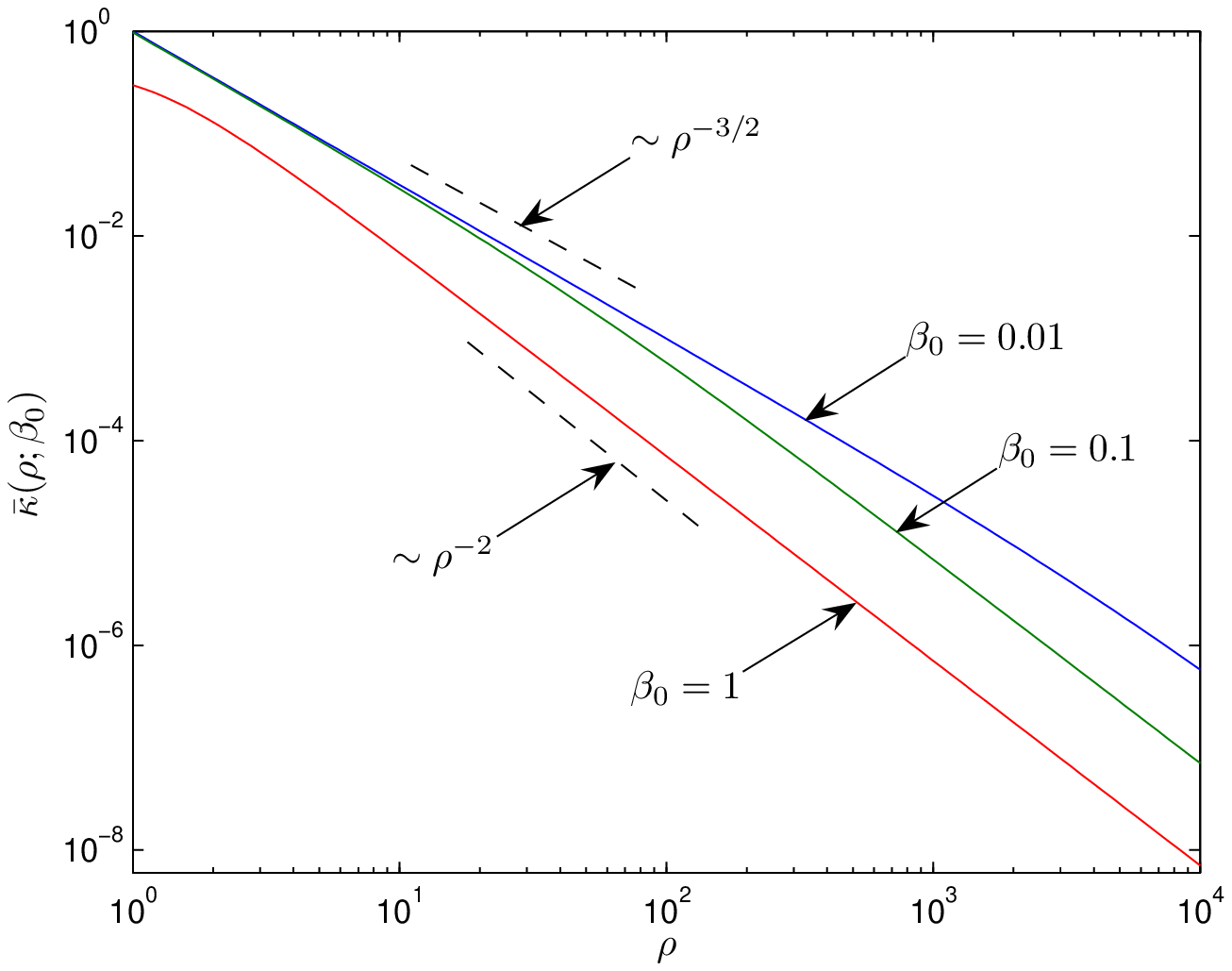}}
{\bf Fig.~5:} \small{Dependence of the normalized pdf $\bar{\kappa}(\rho;\beta_0)$
given by expression \eqref{avepdfkapexpr} of the waiting times
until the completion of the target task as a function of the normalized time $\rho$ for
$\beta=0.01;0.1; 1$. For the smaller values of $\beta$, one can observe the intermediate asymptotic
$\sim \rho^{-3/2}$ progressively crossing over to the tail asymptotic $\sim \rho^{-2}$.}
\end{quote}

\subsubsection{Derivation of the survival distribution $Q(t)$ for a Gaussian and a semi-Gaussian distribution of the normalized time deficit parameter $\beta$ \label{yh3yqefergttnu}}

In the presence of a distribution of time deficit parameters $\beta$, the complementary cumulative
distribution $Q(t)$ of the waiting time until the completion of the target task can be written as
\begin{equation}\label{avcompdist}
\bar{\mathcal{Q}}(\rho;\beta_0) = {1 \over \sqrt{2\pi} \beta_0} \int_{-\infty}^\infty  \mathcal{Q}(\rho;\beta) \exp\left(-{\beta^2 \over 2 \beta_0^2} \right)d \beta~ .
\end{equation}

\begin{quote}
\centerline{
\includegraphics[width=11cm]{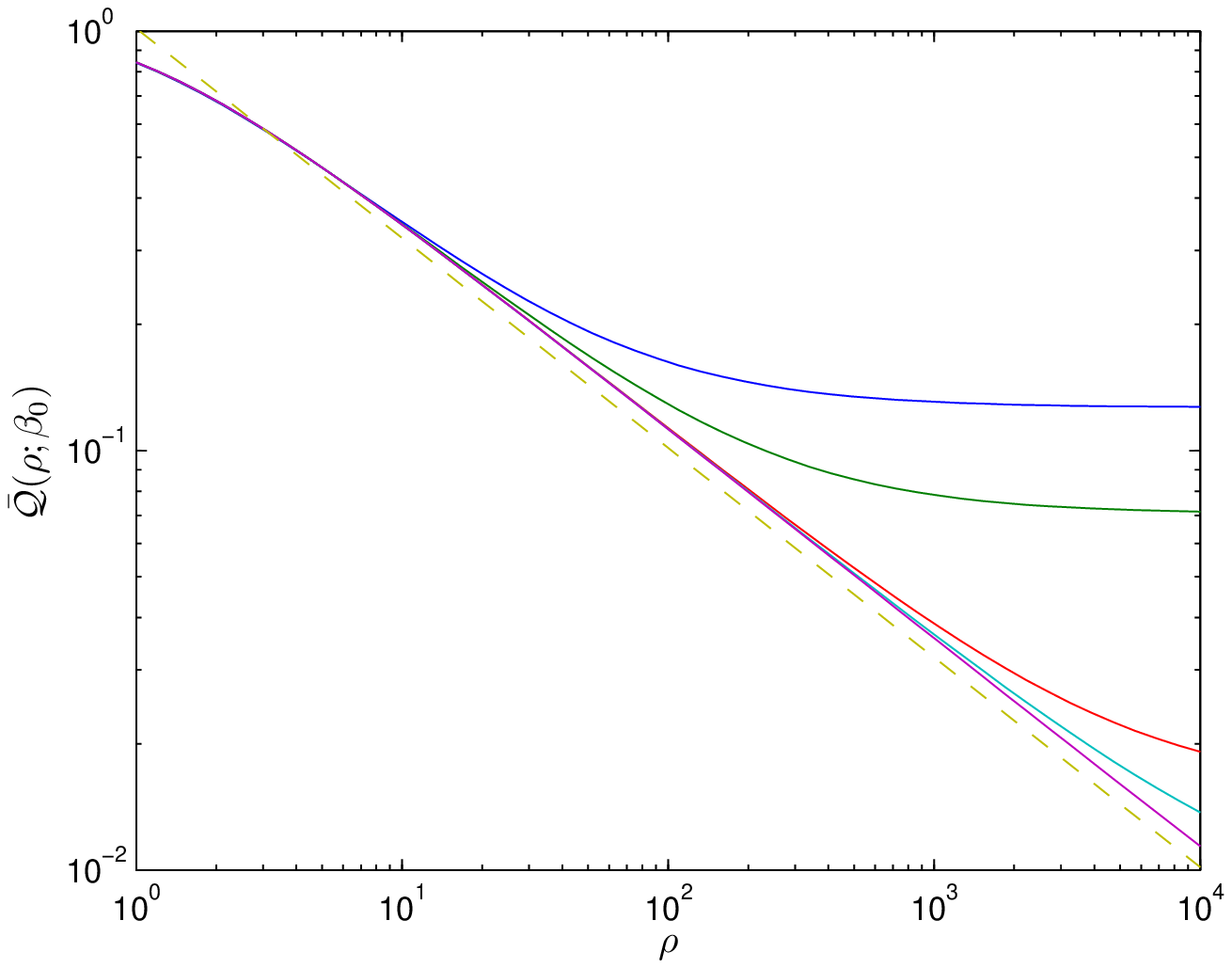}}
{\bf Fig.~6:} \small{Dependence of the averaged cumulative distribution $\bar{\mathcal{Q}}(\rho;\beta_0)$ as a function of the normalized time $\rho$. The solid lines correspond to $\beta_0=0.1;~ 0.05;~0.01; ~ 0.005;~ 0.001$ (top to bottom).
The dashed straight line is the power law asymptotic \eqref{qbetouasymp}.}
\end{quote}

Figure~6 plots of the dependence of $\mathcal{Q}(\rho;\beta_0)$ as a function of $\rho$ for various values
of $\beta_0$. For all non-zero values of $\beta_0$, one can observe that the asymptotic tail exhibits
an upward curvature, leading to a departure from the a priori expected dependence
$\bar{\mathcal{Q}}(\rho;\beta_0)\sim~\rho^{-1}$. And the larger $\beta_0$ is, the slower is the decay of
$\bar{\mathcal{Q}}(\rho;\beta_0)$, which becomes even slower than $\sim \rho^{-1/2}$.

\begin{quote}
\centerline{
\includegraphics[width=11cm]{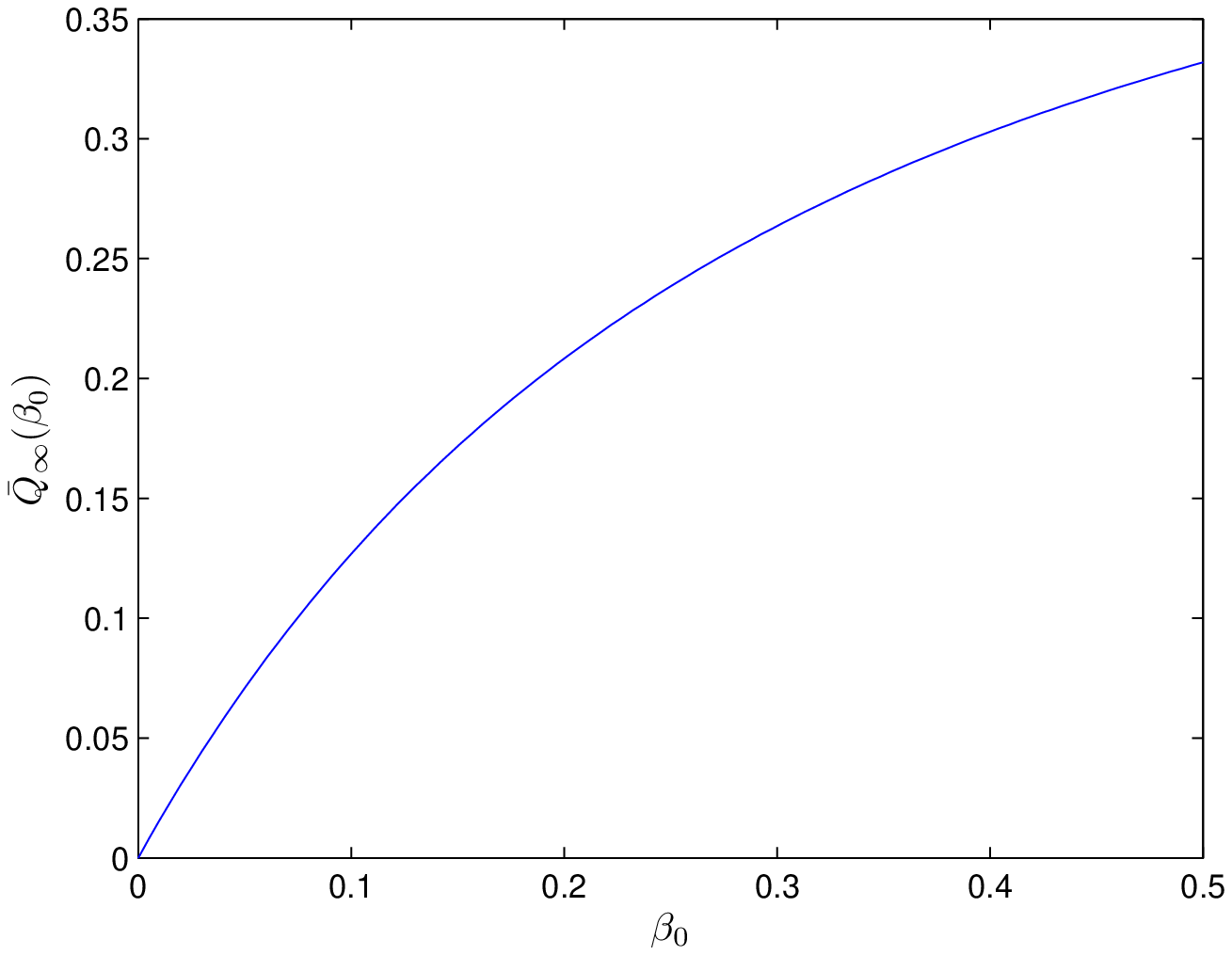}}
{\bf Fig.~7:} \small{Plot of limiting probability $\bar{\mathcal{Q}}_\infty(\beta_0)$, preventing shaping the power law of averaged complementing distribution $\bar{\mathcal{Q}}(\rho;\beta_0)$ for $\rho\to\infty$ }
\end{quote}

The origin of the contradiction between the well-defined asymptotic power law (\ref{avepdfkappowas}) for
the pdf and the behavior shown in figure~6 stems from the existence of the transition occurring
at $\beta_0=0$ above which $ \mathcal{Q}(\rho;\beta)$ acquires a non-zero limit $Q_\infty(\beta)$
given by \eqref{qliminf} at $\rho \to +\infty$, as explained in subsection \ref{thbbns}.
Accordingly, the averaged complementary distribution  $\bar{\mathcal{Q}}(\rho;\beta_0)$
exhibits the strictly positive limit
\begin{equation}\label{barqinf}
\lim_{\rho\to\infty}\bar{\mathcal{Q}}(\rho;\beta_0) = \bar{Q}_\infty(\beta_0)
= \int_0^\infty Q_\infty(\beta) \psi(\beta; \beta_0) d\beta~ .
\end{equation}
For the Gaussian distribution \eqref{gaussbeta}, this limit is given by
\begin{equation}\label{barqinfexpr}
\bar{\mathcal{Q}}_\infty(\beta_0) =  {1 \over 2}\left(1- e^{8\beta_0^2} ~\text{erfc}(2 \sqrt{2} \beta_0)\right)~ .
\end{equation}
Figure~7 shows the dependence of this limit $\bar{\mathcal{Q}}_\infty(\beta_0)$ as a function of $\beta_0$.
In the context of browser updating and software patching, this predicts a regime
in which an intermediate asymptotic $\bar{\mathcal{Q}}(\rho;\beta_0)\sim~\rho^{-1}$ is followed
by a slow cross-over to a positive plateau, corresponding to a finite fraction of the
population that never upgrades or patches.

In subsection \ref{gbja'awqfkv}, we argued that
individuals confronted with a flow rate $\sim 1/\langle \tau \rangle$ of tasks
and the desire to solve them characterized by the average solution time $\langle \eta \rangle$
tend to adjust $\langle \tau \rangle$ towards $\langle \eta \rangle$ and/or vice-versa.
Let us here consider the possibility that $\langle \tau \rangle$ remains marginally
smaller than $\langle \eta \rangle$, so that tasks do not accumulate. Taking into account
the heterogeneity of humans and the variability with time of their strategy, this
corresponds to changing the Gaussian distribution \eqref{gaussbeta}
into the semi-Gaussian distribution\begin{equation}\label{gaussbetalhs}
\psi_-(\beta;\beta_0) =
\begin{cases}
0 ~ , & \beta> 0~ , \\[3mm] \displaystyle
\sqrt{{2\over \pi}} {1 \over\beta_0} \exp\left(-{\beta^2 \over 2\beta_0^2} \right) ~ , & \beta<0~ .
\end{cases}
\end{equation}
The corresponding average complementary cumulative distribution of the waiting times till the completion
of the target task reads
\begin{equation}\label{avcompdistlhs}
\bar{\mathcal{Q}}_-(\rho;\beta_0) =  \sqrt{{2 \over \pi}}{1 \over \beta_0} \int_{-\infty}^0  \mathcal{Q}(\rho;\beta) \exp\left(-{\beta^2 \over 2 \beta_0^2} \right)d \beta~ .
\end{equation}
Since the contributions of non-zero values of $\bar{\mathcal{Q}}_\infty(\beta_0)$ are removed
by this specification of the distribution of $\beta$, $\bar{\mathcal{Q}}_-(\rho;\beta_0)$ exhibits
a well-defined asymptotic power law $\sim\rho^{-1}$.
Figure~8 shows the function $\bar{\mathcal{Q}}_-(\rho;\beta_0)$ as a function of $\rho$ for different $\beta_0$ values,
and illustrates the crossover from the power law $\sim \rho^{-1/2}$ for $\rho < \beta_0^{-2}$ (condition (\ref{powaswithinbou}))
to $\sim\rho^{-1}$ at large times $\rho$.

\begin{quote}
\centerline{
\includegraphics[width=10cm]{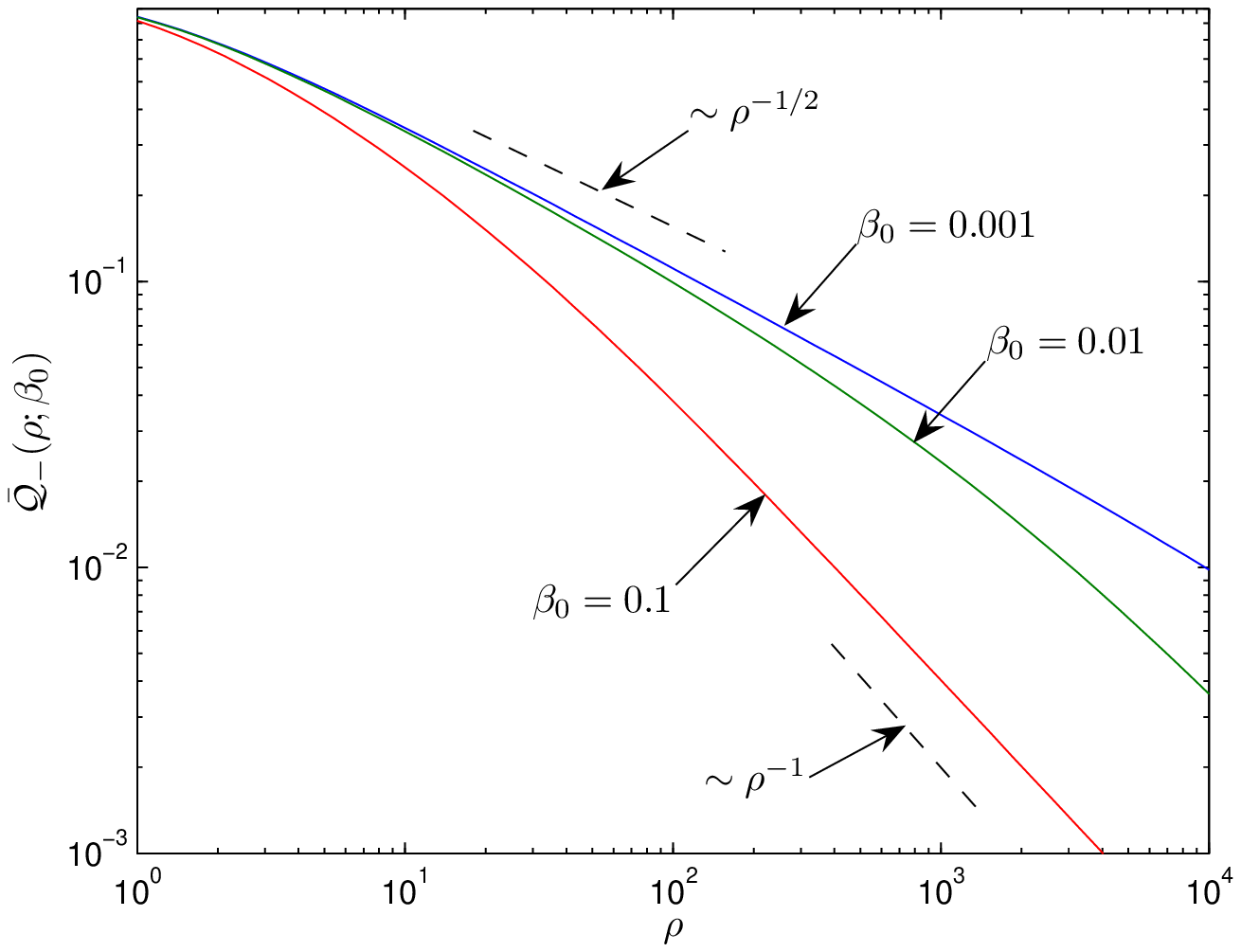}}
{\bf Fig.~8:} \small{Dependence of complementary cumulative distribution $\bar{\mathcal{Q}}_-(\rho;\beta_0)$ given by \eqref{avcompdistlhs} as a function of the normalized waiting time $\rho$ till the completion of the target task. Top to bottom: $\beta_0=0.001; 0.01; 0.1$. One can observe the crossover from $\sim\rho^{-1/2}$ to $\sim\rho^{-1}$ for $\beta_0=0.01$ and
the unique power law $\sim\rho^{-1}$ for $\beta_0=0.1$}
\end{quote}

\subsection{Non-regular distribution of the normalized time deficit parameter $\beta$ around the origin \label{tuhfvnghihd}}

The assumption of a Gaussian or semi-Gaussian pdf (\ref{gaussbeta}) for the
distribution of the normalized time deficit parameter $\beta$ is representative of
the general class of distributions which are regular close to the origin, i.e.,
converge to a non-zero constant for $\beta \to 0$. As we showed
in the previous subsections, it is the regular behavior around $\beta=0$
which controls the tail of the pdf and survival distribution of waiting times.
It is therefore interesting
to investigate the consequence of the existence of less or more probable deviations from
the critical value $\beta=0$. For instance, as mentioned in subsection \ref{gbja'awqfkv},
if individuals adjust their time deficit parameter $\beta \propto  \langle \eta \rangle - \langle \tau \rangle$
via state-dependent control actions, we can expect deviations from the
assumption that the fluctuations of $\beta$ are smooth around $\beta=0$. We
capture such possibility by considering the following asymptotic behavior
of the distribution of $\beta$'s
\begin{equation}\label{attracasympt}
\psi_-(\beta;\beta_0|\nu)\approx A(\beta_0,\nu) |\beta|^\nu~ , \qquad  -1 < \nu < +\infty ~, \qquad \beta\to -0 ~ , \qquad (A<\infty) ~ .
\end{equation}
The case $\nu=0$ recovers the regime of subsection \ref{yhyth3ya}. In order to
remove the impact of the super-critical domain $\beta>0$ on the survival distribution,
we also assume here that all $\beta$'s are negative. It is a simple matter to
remove this condition and recover a regime of non-zero asymptotic behavior
for the survival distribution, as discussed in subsection \ref{yh3yqefergttnu}.

The corresponding pdf of the waiting time till completion of the target task is given by
\begin{equation}\label{pdflhsexact}
\bar{\kappa}(\rho;\beta_0|\nu) = \int_{-\infty}^0 \kappa(\rho;\beta)\psi_-(\beta;\beta_0|\nu) d\beta~.
\end{equation}
In order to determine its asymptotic behavior for large $\rho$'s, it is sufficient to replace this
expression by
\begin{equation}\label{pdflhsapprox}
\bar{\kappa}(\rho;\beta_0|\nu) \approx A(\beta_0,\nu)\int_{-\infty}^0 \kappa(\rho;\beta)|\beta|^\nu d\beta~ ,
\end{equation}
which, using \eqref{kaprhoexpasymp},  yields
\begin{equation}\label{barkappowasymp}
\bar{\kappa}(\rho;\beta_0|\nu) \approx A(\beta_0,\nu) {1 \over 2} \Gamma\left({1+\nu \over 2}\right) \rho^{-\alpha(\nu)-1}~,
 \qquad  \alpha(\nu) = 1+{\nu \over 2} ~, \qquad \rho\to\infty~ .
\end{equation}
Similarly, the survival distribution of waiting times is given by
\begin{equation}\label{compdistlhsexact}
\bar{\mathcal{Q}}(\rho;\beta_0|\nu) = \int_{-\infty}^0 \mathcal{Q}(\rho;\beta)\psi_-(\beta;\beta_0|\nu) d\beta
\sim  \rho^{-\alpha(\nu)}~.
\end{equation}
Since $\nu\in(-1,\infty)$ for $\psi_-(\beta;\beta_0|\nu)$ to be normalized, formula
\eqref{attracasympt} shows that the exponent $\alpha$ can only take values in the interval
\begin{equation}
\alpha\in\left({1 \over 2},\infty\right)~.
\end{equation}
For instance, D\"ubendorfer et al. \cite{Frei_powerlaw}  report a value $\alpha=2/3$ for the
decay of the fraction of computers that still keep an outdated Firefox 2 browser. Within the present
framework, this corresponds to $\nu=-2/3$, i.e., to a significantly stronger concentration of
$\beta$ close to $0$ than would be found expected from a semi-Gaussian distribution for instance.

\section{Theoretical formulation of the impact of procrastination: new power law regimes \label{bwglvlw}}

In the previous sections, we have assumed that, as soon as all other tasks are solved, the individual
addresses without delay the target task with the lowest priority that now comes to the front. In
the present section, we explore
the consequences of the different possibility that procrastination kicks in, so that the target task
is postponed and delayed needlessly due to carelessness or laziness, or for whatever other reason.

\subsection{Model and mathematical solution}

Consider the flow of new tasks occurring at the times given by (\ref{tonen})
and the process $V(k)$ defined in (\ref{vkdef}). Let us denote by $t_n$
the times when $V(k)$ touches the value $-\eta_0'$ from above  (see figure~2),
at which the individual is freed from all tasks except the final target task.
The set $\{t_n\}$ are the beginnings of the time intervals  in which the individual
is free to address the target task.
Let $N(t)$ be the random number of such free moments in the time interval $(0,t)$
and let us call
\begin{equation}\label{prnteqndef}
P(n;t) = \Pr\{N(t)= n\}
\end{equation}
the probability that the number of spare times  in $(0,t)$ is exactly equal to $n$.
We assume that the individual will procrastinate in such a free moment
with probability $0 \leq z < 1$. This is the probability for not upgrading your browser
or non patching your software in one of your free times. For simplicity, we consider $z$
to be independent of the duration of the free time interval. It would not be difficult
to consider alternative specifications dependent on the duration for the free time interval
but, for most reasonable versions, the main result on the power law tails obtained below are not modified.
Assuming that procrastination is independent in successive free time intervals,
the probability that the individual does not complete the target task until time $t$ is given by
\begin{equation}\label{complwaitimedistr}
Q(t,z) = \Pr\{\mathcal{T}>t\} = \sum_{n=0}^\infty P(n;t) z^n ~ ,
\end{equation}
where $\mathcal{T}$ is the waiting time till the target task is completed.

In order to calculate $Q(t,z)$ given by \eqref{complwaitimedistr}, we need the expression of $P(n;t)$.
For this, we relate it to the probability
\begin{equation}\label{ftndef}
F(t;n) = \Pr\{t_n<t\}
\end{equation}
that, for a given $n$, the random variable $t_n$ does not exceed $t$. The relation
between $P(n;t)$ and $F(t;n)$ is
\begin{equation}\label{pthrufrel}
P(n;t) = F(t;n)-F(t; n+1) \quad (n\geqslant 1)~, \qquad P(0;t) = 1- F(t;1)~ .
\end{equation}
Expression (\ref{pthrufrel}) writes that the number $n$ of free intervals occurring in $(0,t)$
is determined by the condition that the $n$-th free time interval starts before $t$
while the $(n+1)$-th free time interval starts after $t$.

Substituting (\ref{pthrufrel}) in relation \eqref{complwaitimedistr} yields
\begin{equation}\label{qtnsumfs}
Q(t,z) = 1+ (z-1) \sum_{n=1}^\infty F(t;n) z^{n-1}~ .
\end{equation}
It is more convenient to work with the pdf of the waiting time till the completion of the target task,
defined by $q(t,z) = - {\partial Q(t,z) \over \partial t}$. From (\ref{qtnsumfs}), we obtain
\begin{equation}\label{qtzsumpdfs}
q(t,z) = (1-z) \sum_{n=1}^\infty f(t;n) z^{n-1}~ ,
\end{equation}
where
\begin{equation}
f(t;n) = {\partial F(t;n) \over \partial t}
\end{equation}
is the pdf of the random variable $t_n$ of the beginning of the $n$-th free time interval.

As it should, the limit $z=0$ in \eqref{qtzsumpdfs} recovers the pdf given by (\ref{qnt5}) of the waiting time till the completion of the target tasks,
\begin{equation}
q(t,0) = f(t;1) \equiv f_1(t)~.
\end{equation}

The beginning time $t_n$ of the $n$-th free interval can be written as the sum of $n$
waiting times:
\begin{equation}
t_n = \Delta t_1 + ...+ \Delta t_n~.
\label{tjbqhxci}
\end{equation}
The first waiting time $\Delta t_1$ is the duration of the time interval starting at the inception time $t=0$ of the target task
until the individual is free to address the target task for the first time. We denote its pdf as $f_1(t)$.
The other terms $\Delta t_2, ..., \Delta t_n$ quantify the waiting times between successive
beginnings of free time intervals.  They are independent identically distributed
random variables, with common pdf denoted $f(t)$. The expressions for $f_1(t)$ and $f(t)$
are made explicit in subsection \ref{tiqyyyww}. Then, the pdf of the sum \eqref{tjbqhxci} is equal to the $n$-times convolution
\begin{equation}\label{fntimesconv}
f(t;n) = f_1(t)\otimes \underbrace{f(t)\otimes \cdots \otimes f(t)}_{n-1 ~ \text{times}}
\end{equation}
The Laplace transform of $f(t;n)$ is thus
\begin{equation}
\hat{f}(s;n) \equiv  \int_0^\infty f(t;n) e^{-s t} dt =  \hat{f}_1(s) \cdot \hat{f}^{n-1}(s)~ ,
\end{equation}
where $\hat{f}_1(s)$ and $\hat{f}(s)$ are respectively the Laplace transforms of $f_1(t)$ and $f(t)$.

Applying the Laplace transform to both sides of equality \eqref{qtzsumpdfs}, we obtain
\begin{equation}\label{qtnsumfslap}
\hat{q}(s,z) =  (1-z) \hat{f}_1(s) \sum_{n=1}^\infty \hat{f}^{n-1}(s) z^{n-1} =
{(1-z)\hat{f}_1(s) \over 1 - z \hat{f}(z)} ~ .
\end{equation}
As shown in subsection \ref{tiqyyyww}, the pdf's $f_1(t)$ and $f(t)$ have
in general the following power law asymptotic
\begin{equation}\label{ffoneasymp}
f_1(t) \simeq a_1 \cdot t^{-\alpha-1}~ , \qquad f(t) \simeq a \cdot t^{-\alpha-1}~ , \qquad t\to \infty~ .
\end{equation}
This implies that their Laplace transform have the following asymptotic
\begin{equation}\label{lapimffoneasymp}
\hat{f}_1(s) \simeq 1 + \Gamma(-\alpha) \cdot a_1 \cdot s^\alpha~ , \qquad
\hat{f}(s) \simeq 1 + \Gamma(-\alpha) \cdot a \cdot s^\alpha~ , \qquad s\to 0~ .
\end{equation}
Substituting these last relations into \eqref{qtnsumfslap} yields the asymptotic form of $\hat{q}(s,z)$
\begin{equation}\label{qompllaplasymp}
\hat{q}(s,z) \simeq {1 \over 1- \chi s^\alpha}~ , \qquad s\to 0~ ,
\end{equation}
where
\begin{equation}
\chi = - a\cdot \Gamma(-\alpha) \cdot {z \over 1-z} \qquad (\chi>0)~ .
\label{tjjvqelv,q}
\end{equation}

\subsection{Expression of the probability $z$ for not completing the target task in one of the free times}

Let us denote by $\zeta_k$ the duration of the $k$-th free interval, from the beginning time $t_k$ to the
arrival of the first new task. Note that one cannot interpret $\zeta_k$ as the duration of the time interval
during which the Wiener process $V(t)$ remains below the level $-\eta_0'$, because
the formulation in terms of a Wiener process has a sense only for $V(t)>-\eta_0'$.
Actually, $\zeta_k$ has the simple interpretation of being the waiting time counted
from any arbitrary time till the occurrence of a new task.

Consider the simple instance in which the tasks arrive according to a Poisson flow with rate $\lambda$,
such that the pdf of $\zeta_k$ reads
\begin{equation}\label{pizetexp}
\pi(\zeta) = \lambda e^{-\lambda \zeta}~ .
\end{equation}
One can interpret $1/\lambda$ as the mean waiting time between task arrivals.
For the sake of clarity, let us also assume that the probability not to perform the target task during a free time
is a decreasing exponential function of the duration $\zeta_k$ of that free interval:
\begin{equation}
P(\zeta) = e^{-\lambda_1 \zeta}~ .
\label{yhjqq,a,}
\end{equation}
One can interpret $1/\lambda_1$ as the average ``procrastination time.''
Expression (\ref{yhjqq,a,}) assumes that the individual decides to perform the target task
during one of her free time according to a constant probability per unit time, i.e., according to another
Poisson process with rate $\lambda_1$.
Averaging this probability over the statistics of $\zeta$ yields the probability $z$ that the target task
will not be performed during a given free time interval:
\begin{equation}\label{zcalc}
z = \int_0^\infty P(\zeta) \pi(\zeta) d\zeta = \int_0^\infty e^{-\lambda_1 \zeta} \lambda e^{-\lambda \zeta} dz = {\lambda \over \lambda+ \lambda_1}~ .
\end{equation}
If $\lambda_1 \ll \lambda$, i.e., if the arrival rate of new tasks is significantly larger than the rate with which
the individual fights her procrastination, then the probability $z$ not to perform
the target task in a given free interval is close to unity. As we see below, this regime $z \to 1$
is responsible for a much slower decay of the pdf and survival distribution of the waiting times
till the completion of the target task.

\subsection{Derivation of the distribution of waiting times till the completion of the target tasks
in the presence of pronounced procrastination ($z \to 1$)}

It is well-known (see, for instance, \cite{SZ,PSW}) that the inverse Laplace transform of \eqref{qompllaplasymp} is equal to
\begin{equation}\label{qtalphmod}
q(t,z) = {1 \over \chi^{1/\alpha}} \kappa\left({t \over \chi^{1/\alpha}}, \alpha\right)~ ,
\end{equation}
where $\kappa(y,\alpha)$ can be expressed as the weighted sum of exponential distributions
\begin{equation}\label{kappaintexp}
\kappa(y,\alpha) = \int_0^\infty {1 \over \mu} \exp\left(-{y \over \mu} \right) \zeta(\mu,\alpha) d\mu~ ,
\end{equation}
with weights
\begin{equation}\label{zetaexpr}
\zeta(\mu,\alpha) = {1 \over \pi \mu} {\sin(\pi \alpha) \over \mu^\alpha + \mu^{-\alpha} + 2 \cos(\pi \alpha)} ~ .
\end{equation}
The corresponding complementary distribution is
\begin{equation}\label{komplrenormdist}
\mathcal{K}(y,\alpha) \equiv  \int_{-\infty} ^y \kappa(x,\alpha) dx = \int^{\infty}_0\exp\left(-{y \over \mu} \right) \zeta(\mu,\alpha) d\mu~ .
\end{equation}
Expression (\ref{kappaintexp}) predicts the existence of the two regimes
\begin{equation}
\kappa(y,\alpha) \sim {1 \over y^{1-\alpha}}~,~{\rm for}~y \ll 1~;
\qquad \text{and} \qquad \kappa(y,\alpha) \sim  {1 \over y^{1+\alpha}}~,~{\rm for}~y \gg 1~.
\label{yhjkqmfdl,oo}
\end{equation}
This translates into
\begin{equation}
q(t, z) \sim \chi^{-1}~ {1 \over t^{1-\alpha}}~,~{\rm for}~t \ll \chi^{1 /\alpha}~;
\qquad \text{and} \qquad  q(t, z) \sim \chi ~ {1 \over t^{1+\alpha}}~,~{\rm for}~t \gg \chi^{1 /\alpha}~.
\end{equation}
Figure~9 plots the pdf $\kappa(y,\alpha)$ given by \eqref{kappaintexp}  for $\alpha=0.7$
and confirms the existence of an intermediate asymptotic power law regime
$q(t, z) \sim  1/ t^{1-\alpha}$ for $t \ll \chi^{1 /\alpha}$, which decays much slower than in absence of
procrastination ($z=0$ leading to a small $\chi$ and to $q(t, z) \sim  1/ t^{1+\alpha}$). The two regimes
could be observed in the data by taking the derivative as a function of time of
the fraction of remaining individuals who have not yet solved the target task.

For $\lambda_1 \ll \lambda$, we can approximate expression (\ref{zcalc})
by $z=1-{\lambda_1 \over \lambda}$, so that $\chi$ given by (\ref{tjjvqelv,q}) is
approximately equal to $\chi = - a\cdot \Gamma(-\alpha) \cdot {\lambda \over \lambda_1} \gg 1$, i.e., it is
proportional to the ratio of the average ``procrastination time'' over the mean waiting time between task arrivals.
It is this ratio ${\lambda \over \lambda_1}$ that determines the range of the intermediate
asymptotic power law $q(t, z) \sim  1/ t^{1-\alpha}$, which holds for $t \ll (\lambda / \lambda_1)^{1 /\alpha}$.

\begin{quote}
\centerline{
\includegraphics[width=13cm]{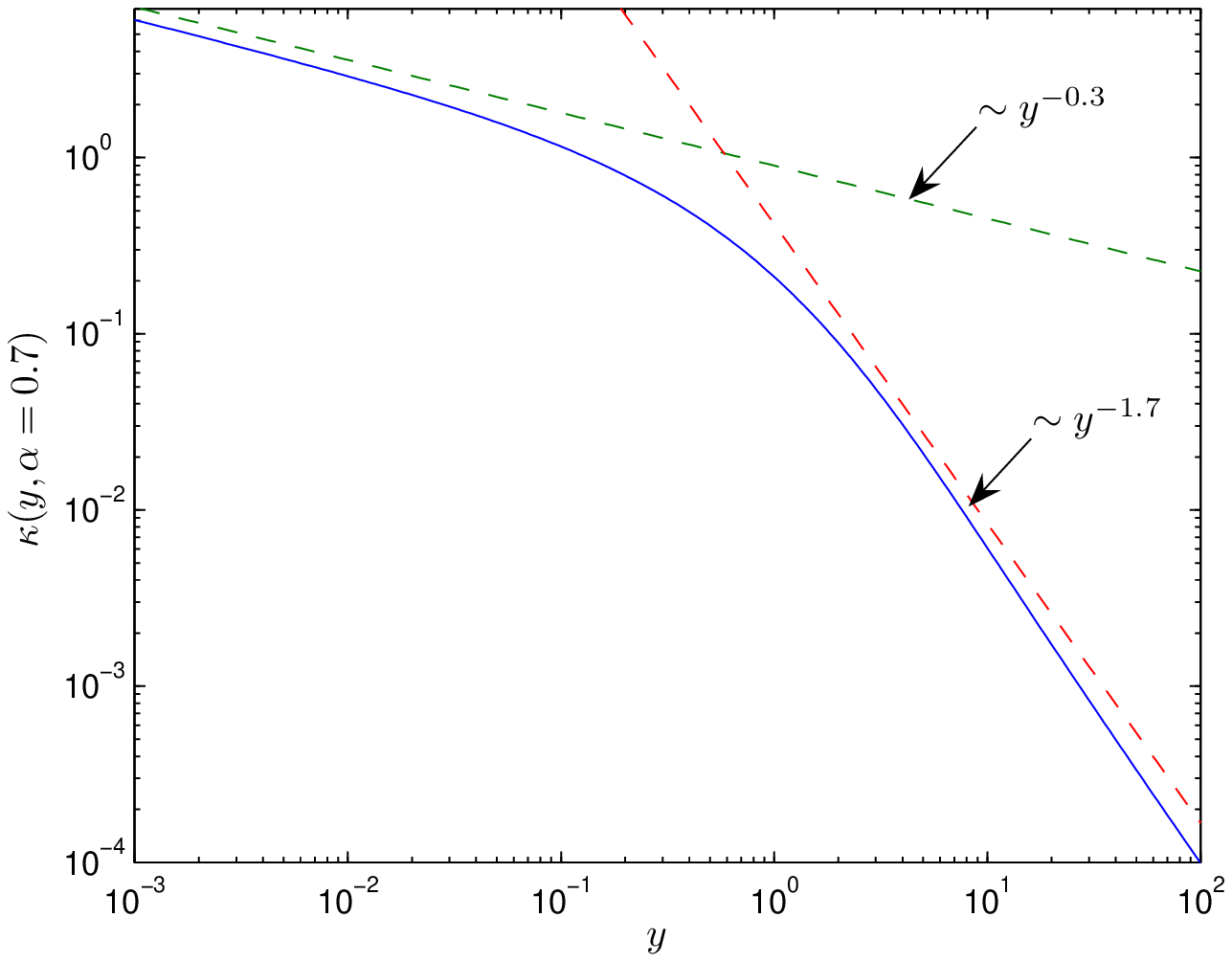}}
{\bf Fig.~9:} \small{Plot of the pdf $\kappa(y,\alpha)$ given by \eqref{kappaintexp} as a function of the reduced
variable $y=t/\chi^{1/\alpha}$ for $\alpha=0.7$. The two dashed lines correspond to the intermediate
asymptotic and to the tail asymptotic given by (\ref{yhjkqmfdl,oo}).}
\end{quote}

If we look directly at this fraction ($\mathcal{K}(y,\alpha)$ in normalized units) of non-solved target task,
we do not find two clear power law regimes, but rather a smooth cross-over to the asymptotic
power law tail $\mathcal{K}(y,\alpha) \sim 1/y^{\alpha}$, as shown in figure~10.

\begin{quote}
\centerline{
\includegraphics[width=13cm]{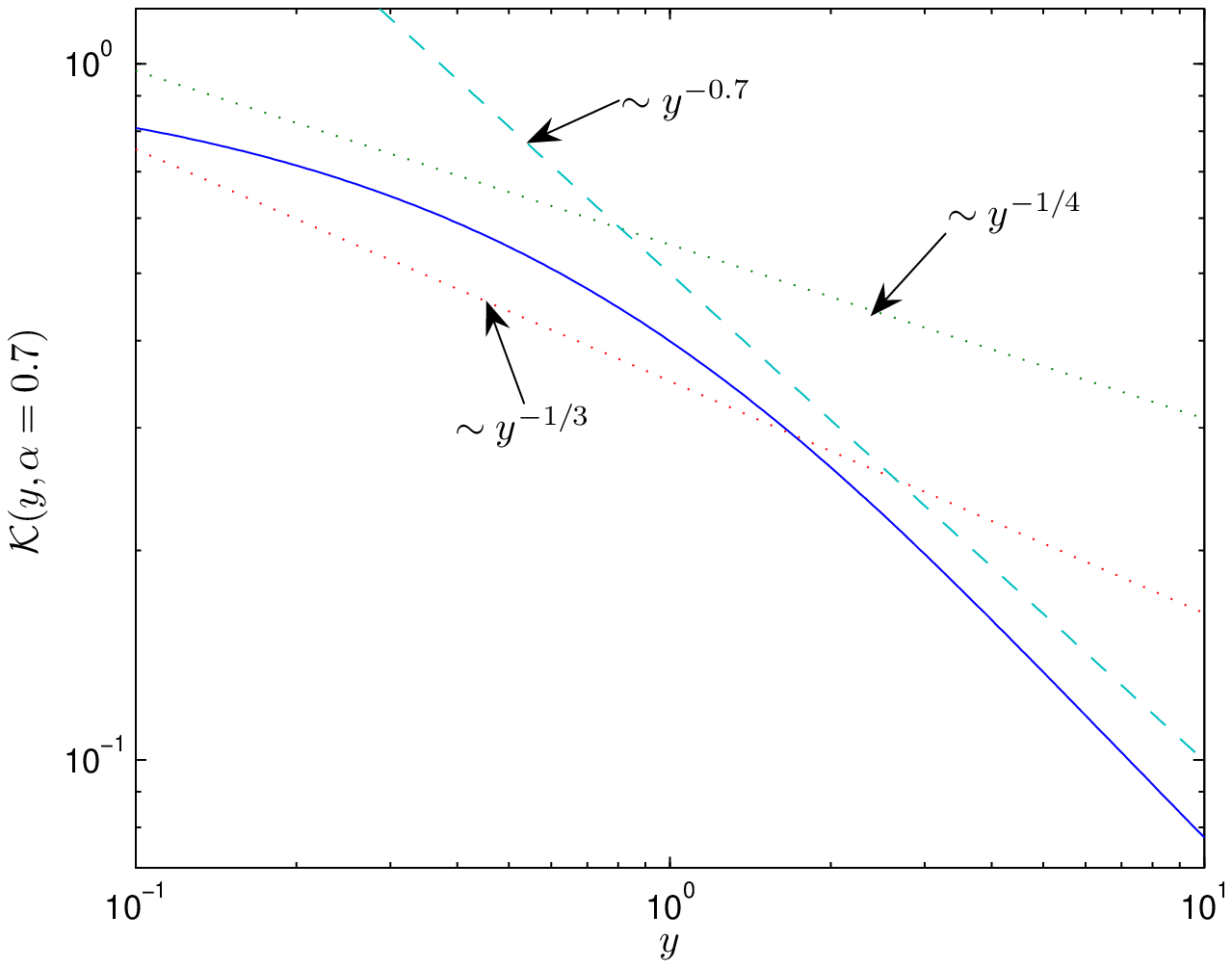}}
{\bf Fig.~10:} \small{Dependence of the normalized complementary cumulative distribution $\mathcal{K}(y,\alpha)$ given by \eqref{komplrenormdist}, for $\alpha=0.7$, as a function of the normalized time $y=t/\chi^{1/\alpha}$. The dashed straight line corresponds to the asymptotic power law $\sim y^{-\alpha}$. The two dotted straight lines are power laws $\sim y^{-1/3\div 1/4}$, to help the eye suggest the presence of an intermediate apparent slow power law decay over some
limited time range.}
\end{quote}

\subsection{Derivation of the pdf of the waiting times $\Delta t_k$ between successive
beginnings of free time intervals \label{tiqyyyww}}

Let us now justify the form (\ref{ffoneasymp}) for the pdf $f_1(t)$ of $\Delta t_1$ and of the pdf $f(t)$
of the other independent random variables $\Delta t_2, ..., \Delta t_n$ as defined in (\ref{tjbqhxci}).

As we showed in section \ref{jguuuurrw},
the pdf of $\Delta t_1$ coincides (in the Wiener process approximation) with the pdf
for the Wiener process $V(t)$ of first touching the level $-\eta_0'$:
\begin{equation}\label{fonefirsttouch}
f_1(t) = {\gamma \over \left<\tau\right> \sqrt{2\pi} \theta^{3/2}} \exp\left(-{(\delta \theta + \gamma)^2 \over 2 \theta}\right) ~ , \qquad \theta = {t \over \left<\tau\right>} ~ .
\end{equation}
This has the form (\ref{ffoneasymp}) with $\alpha=1/2$, when the time deficit
parameter $\delta$ is close to zero.

The other random variables $\Delta t_2, ..., \Delta t_n$ are each the sum of two independent contributions,
$\Delta t_k = \zeta_k + {\hat \Delta t}_k $,  where
(i) $\zeta_k$ is the  duration of a free time interval, which has the same distribution as that of the waiting time counted from any arbitrary time till the occurrence of a new task and (ii) ${\hat \Delta t}_k $ is a time
similar to $\Delta t_1$ for the Wiener process $V(t)$ of first touching some new level.
In order to specify further the properties of this second contribution ${\hat \Delta t}_k $, we recall that the
pdf $f_1(t)$ of  $\Delta t_1$ given by \eqref{fonefirsttouch} depends on the parameter
$\gamma = {\eta' \over \sigma_\eta}$ defined in (\ref{taudelnot}),
where $\eta_0'$ is the time needed by the individual to solve tasks
that has been stored, and $\sigma_\eta$ is the standard deviations of the times
$\eta_k$ needed to solve the $k$-th task. The pdf ${\hat f}(t)$ of ${\hat \Delta t}_k$ can be written as
\begin{equation}\label{ftgenrel}
{\hat f}(t) = \int_0^\infty w(\gamma) f_1(t|\gamma) d\gamma~ ,
\end{equation}
where the notation $f_1(t|\gamma)$ makes explicit the dependence on $\gamma$ in expression (\ref{fonefirsttouch}).
The integral in (\ref{ftgenrel}) is performed over the random variable
$\gamma$ weighted by its pdf $w(\gamma)$, which is determined as follows.
The end of an interval which was free of any task (except the target task that remains
to be addressed) is triggered by the occurrence of a new task that takes priority over the target task,
and the pdf of the time needed to solve it is $\varphi(\eta)$, with mean and variance
equal to  $\left<\eta\right>$ and $\sigma_\eta^2$, as defined in subsection \ref{rbk'r}.
We thus have
\begin{equation}
w(\gamma) = \sigma_\eta \varphi\left( \sigma_\eta \gamma\right)~ .
\end{equation}
To illustrate, suppose that $\varphi(\eta)$ is exponential, $\varphi(\eta) = \mu e^{-\mu \eta}$,
leading to $\sigma_\eta= 1/\mu$ and
\begin{equation}
w(\gamma) = e^{-\gamma}~.
\label{wgammexpr}
\end{equation}
Substituting \eqref{fonefirsttouch} and \eqref{wgammexpr} into \eqref{ftgenrel} yields
\begin{equation}
{\hat f}(t)= {1 \over 2 \left<\tau\right>}
\exp\left(- {\delta^2 \theta \over 2} \right) \left( \sqrt{{2\over \pi\theta}} - \exp\left(- {(1+\delta)^2 \theta \over 2}\right)(1+\delta) \text{erfc}\left({(1+\delta) \sqrt{\theta} \over \sqrt{2}}\right)\right)~ .
\end{equation}
It is straightforward to check that this pdf ${\hat f}(t)$ has the asymptotic power law (\ref{ffoneasymp}) with $\alpha=1/2$.
Now, $f(t)$ is the convolution of the pdf $\pi(\zeta)$ and of ${\hat f}(t)$, and its tail is determined
by that of ${\hat f}(t)$, hence the form (\ref{ffoneasymp}) with $\alpha=1/2$.

\section{Concluding remarks \label{tghtgta}}

We have developed a simple general framework to model the distribution of
waiting times between the triggering factor and the actual
realization of a job, for the particular tasks that are both important
(sometimes even essential) but are often considered low priority
because they require interrupting the normal flow of work or life. Beyond
the examples of Internet browser updates and software vulnerability
patching which initially motivated our interest in this question,
we suggest that our theory can apply to less quantifiable but equally important
questions such as the delay in implementing important decisions
in one's life. While we recognize of course the existence of additional
important psychological factors and social influences, our approach
provides a simple parsimonious starting point of a general theory
of procrastination.

\vskip 1cm
{\bf Acknowledgements}:  We are grateful to R. Crane and T. Maillart for stimulating discussions.


\clearpage

\begin{thebibliography}{widest-label}

\bibitem{Eck} Eckmann, J.-P., E. Moses and D. Sergi,
Proc. Nat. Acad. Sci. USA, 101(40), 14333-14337 (2004).

\bibitem{Barabasi_Nature05} A.-L. Barab\'asi, Nature 435, 207 (2005).

\bibitem{Oliveira_Bara} Oliveira, J.G. and A.-L. Barab\'asi,
Nature  437, 1251 (2005).

\bibitem{Vasquez_et_al_06} Vazquez, A., J. G. Oliveira, Z. Dezso, K. I. Goh, I. Kondor, and A. L. Barabasi,
Physical Review E 73, 036127 (2006).

\bibitem{roehner2004news} Roehner, M., Sornette, D. and J. V. Andersen,
Int. J. Mod. Phys. C 15 (6), 809-834 (2004).

\bibitem{johansen2000internaut} Johansen, A. and D. Sornette,
Physica A 276, (1-2), 338-345 (2000).

\bibitem{johansen2001internaut} Johansen, A.,
Physica A 296 (3-4), 539-546 (2001).

\bibitem{sornette2004amazon} Sornette, D., F. Deschatres, T. Gilbert, and Y. Ageon,
Test Using Book Sale Ranking, Phys. Rev. Lett. 93, 228701 (2004).

\bibitem{deschatres2005amazon} Deschatres, F. and D. Sornette,
Phys. Rev. E 72, 016112 (2005).

\bibitem{cranesorYouTube} Crane, R. and D. Sornette,
Proc. Nat. Acad. Sci. USA 105 (41), 15649-15653 (2008).

\bibitem{sornette2005origins} Sornette, D. (2005) in \textit{Extreme
Events in Nature and Society}, eds Albeverio S, Jentsch V, Kantz H
(Springer, Berlin), pp 95-119.

\bibitem{CraneSchSor_donation} Crane, R.,  F. Schweitzer and D. Sornette,
New Power Law Signature of Media Exposure in Human Response Waiting Time Distributions,
submitted to Physical Review E (2009)
(http://arxiv.org/abs/0903.1406)

\bibitem{Grinstein1} Grinstein, G. and R. Linsker,
Phys. Rev. E 77, 012101 (2008).

\bibitem{Grinstein2} Grinstein, G. and R. Linsker, Phys. Rev. Lett. 97, 130201 (2006).

\bibitem{Redner_first} Redner, S., A Guide to First-Passage Processes,
Cambridge University Press (2007).

\bibitem{Frei1} Frei, S., T. D\"ubendorfer, G. Ollmann  and M. May, Examination of vulnerable
onlineWeb browser populations and the ``insecurity iceberg'', ETH Tech Report, (2008).

\bibitem{Frei2} Frei, S., T. D\"ubendorfer, B. Plattner, Firefox (In)Security Update Dynamics Exposed, ACM SIGCOMM Computer Communication Review (2009).

\bibitem{Jonathan} Gysel, J.,
Analysis and Modeling of Internet Epidemics, Master Thesis MA-2008-01, ETH Zurich (june 2008).

\bibitem{Frei_powerlaw}  D\"ubendorfer, T., Frei, S., J. Gysel, T. Maillart and D. Sornette, Power law
of fading danger and success in the Internet world: worm remanence and browser uses,
ETH Zurich preprint (2009)

\bibitem{Cabrera_Milton04} Cabrera, J.L. and J.G. Milton,
Chaos 14(3),  691-698 (2004).

\bibitem{Eurich_Pawelzik05} Eurich, C.W.  and K. Pawelzik, Optimal control yields power-law behavior, in: Artificial Neural Networks: Formal Models and their Applications, Springer Lecture Notes in Computer Science 3697, W. Duch, J. Kacprzyk, E. Oja, and S. Zadrozny (eds.), Springer-Verlag, Berlin, 365-370 (2005).

\bibitem{Eurich_Pawelzik07} Patzelt, F., M. Riegel, U. Ernst and K. Pawelzik,
Frontier in Computational Neuroscience 1 (4), 1-9 (2007).

\bibitem{SweepSor94} Sornette, D.,
J.Phys.I France 4, 209-221 (1994).

\bibitem{SZ} Saichev A. and Zaslavski G. 
Chaos, 7(4), 753-764 (1997). 

\bibitem{PSW} Piryatinska A., Saichev A.I. and Woyczynski W.A.,
Physica A: Statistical Mechanics and Applications, 349, 375-421 (2005).


\end{thebibliography}
\end{document}